\documentclass[seceq]{ptptex}

\usepackage{graphicx}
\usepackage{bm}
\usepackage{amsmath,amsfonts,latexsym,amssymb} 
\usepackage{verbatim,amsthm,graphicx} 
\usepackage{wrapft}

\usepackage{amsmath}
\usepackage{amssymb}
\usepackage{mathrsfs}

\title{
Exact solutions of higher dimensional black holes%
}

\author{
Shinya \textsc{Tomizawa}${}^{*}$ and Hideki \textsc{Ishihara}${}^{**}$ 
}

\inst{
${}^{*}$Cosmophysics Group, Theory Center, Institute of Particle and Nuclear Studies,
KEK, 1-1 Oho, Tsukuba, Ibaraki, 305-0801, Japan \\
${}^{**}$Department of Mathematics and Physics,
Graduate School of Science, Osaka City University,
3-3-138 Sugimoto, Sumiyoshi, Osaka 558-8585, Japan
}

\abst{ We review exact solutions of black holes in higher dimensions, focusing on asymptotically flat black hole solutions and  Kaluza-Klein type black hole solutions.
 We also summarize some properties which such black hole solutions reveal.
}

\begin{document}

\maketitle

\vspace{-8.0cm}
\hfill{KEK-TH 1450 OCU-PHYS 351 AP-GR 92 }
\vspace{7.5cm}

\tableofcontents

\section{Introduction}

The Einstein theory of gravitation, which is described by a certain complicated system of simultaneous non-linear partial derivative equations, are the great success in modern physics, for example, in the fields of astrophysics or cosmology.
In particular, the discoveries of exact solutions describing black holes such as the Schwarzschild solution~\cite{Sch} and the Kerr solution~\cite{Kerr} have so far provided us with a great deal of insight into various gravitational phenomena.
It must have been one of the most exciting predictions to us in general relativity that there may be black holes in our universe.
Of course, in general, one has a lot of purely mathematical and technical problems, when attempting to find exact solutions and to classify all of them, due to the complexity of the Einstein equations, essentially, its non-linearity. 
Nevertheless, as for the exact solutions of black holes, its classification has been completely achieved,  i.e., it is well known that the Kerr solution is the only black hole solution within the pure gravitational theory (so-called uniqueness theorem for black holes).
Hence, there is no room for asking if there exists a new black hole solution within the pure gravity in four dimensions.

\medskip
In recent years, motivated by string theories, various types of black hole solutions in higher dimensions have so far been found, with the help of, in part, recent development of solution generating 
techniques. 
It is now evident that even within the framework of vacuum Einstein gravity, there is a much richer variety of black hole solutions in higher dimensions. 
In fact, as shown by Emparan and Reall, five-dimensional asymptotically flat vacuum Einstein gravity admits the coexistence of a rotating spherical hole and two rotating rings with the same conserved 
charges, illustrating explicitly the nonuniqueness property of black holes in higher dimensions. Therefore, one may expect that a lot of black hole solutions still remain unfound even in the simplest pure gravity in higher dimensions. 

\medskip
In most cases, in higher dimensions, {\sl asymptotically flat} black holes have been considered, in various theories, for stationary, axisymmetric (with multiple rotational symmetries) black holes being non-compact, as simple higher dimensional
generalizations of the well-known four-dimensional black holes. 
However, since our real, observable world is macroscopically four dimensional, then extra dimensions have to be compactified in realistic spacetime models in a certain appropriate way.
Therefore it is of great interest to consider higher dimensional Kaluza-Klein black holes with compact extra dimensions, which behaves as higher dimensions near the horizon but look four-dimensional for an observers at large distance\footnote{
When the black holes will be small enough compared with the size of the extra dimensions, they may be well approximated by the higher dimensional asymptotically flat black hole solutions. }
Though one would expect that the simplest structure of Kaluza-Klein type black holes are direct products of four dimensions and extra dimensions, one hardly construct exact black hole solutions of these types because of a lack of geometrical symmetry. 
However,  a certain class, called cohomogeneity-one, of exact {\sl Kaluza-Klein black hole} solutions can be obtained by the squashing from the same class of non-compactified black hole solutions, e.g., asymptotically rotating black hole solutions with equal angular momenta, or 
non-asymptotically flat G\"odel black holes with closed timelike curves.

\medskip
The main purpose of this chapter is to review and summarize known (at present) exact solutions of black holes in higher dimensional Einstein gravity, in particular, focusing on a class of asymptotically flat black holes and squashed Kaluza-Klein black holes.  
As far as asymptotically flat black holes is concerned, although there is some overlap with the earlier nice review articles~\cite{ER-review,ER-review2}, we here include the detail of the black ring with $S^2$-rotation~\cite{MishimaIguchi,Fig}, black lens~\cite{Evslin,CT} and multiple black holes~\cite{Elvang-Fig,diring,Elvang-R,Izumi,TanTeo,multi-MP}.
As for a class of Kaluza-Klein black holes, based on our earlier several works, we include not only black holes in the five-dimensional Einstein gravity but also  charged black holes (of bosonic sector) in the five-dimensional minimal supergravity, namely, 
the five-dimensional Einstein-Maxwell-Chern-Simons theory. 
We in detail review black hole solutions obtained from such the squashing and also summary all the solutions found  at present.
The supersymmetric black holes or black hole with a cosmological constant are beyond description of this chapter.

\medskip
This review is organized as follows:
In the following section, %\ref{sec:AF}
we will review asymptotically flat black holes solutions in $D\ (\ge 5)$-dimensional Einstein theories (in the standard sense that a spacetime asymptotes to a Minkowski spacetime at infinity), which include the Schwarzschild-Tangherlini solution (subsection~\ref{sec:Sch}), Myers-Perry solution (subsection~\ref{sec:MP}). 
In subsection~\ref{sec:SA}, we review stationary and axisymmetric solutions with $D-2$ commuting Killing vector fields and the concept of rod structure.
In subsection~\ref{sec:5D}, we review all known five-dimensional asymptotically flat black hole solutions such that the spatial cross section of the horizon is topologically non-spherical, 
namely, black ring solutions, black lens solution, black saturn solution, black di-ring solution and bicycling black ring solution.
Furthermore, we devote  section~\ref{sec:KK} to Kaluza-Klein black hole solutions in five dimensions.
In subsection~\ref{sec:rasheed} we mention the most general Kaluza-Klein black hole solution in the five-dimensional Einstein theory.
Then, in subsection~\ref{sec:IM}, we focus on charged Kaluza-Klein black hole solutions in the bosonic part of the five-dimensional supergravity
which can be obtained by the deformation of the so-called squashing transformation.

%\subsection*{Notations and Conventions}
%In this chapter we use capital Latin indices $M,N,\cdots(=0,1,\cdots,D-1)$ to denote space or time components of a tensor in higher dimensions.
%Greek indices $\mu,\nu,\cdots(= 0,1,2,3)$ are used to denote spacetime components in four dimensions.

\section{Asymptotically flat black holes}\label{sec:AF}
In this section, we consider the exact solutions of {\it asymptotically flat}~\cite{Myers:1986un}, stationary black holes in the theories describing by the $D$-dimensional Einstein-Hilbert action, 
which is written in the form
\begin{eqnarray}
S=\frac{1}{16\pi G} \int d^Dx\sqrt{-g}\ R,
\end{eqnarray}
where $G$ is a Newton's constant in $D$-dimensions. 
We here mean by asymptotically flat that a spacetime asymptotes to $D$-dimensional Minkowski spacetime at infinity, namely,  in terms of Cartesian coordinates $\{t,x^i\}\ (i=1,\cdots,D-1)$ (under a certain gauge condition~\cite{Myers:1986un}), the spacetime metric at infinity,
$r:=\sqrt{\sum_{i=1}^{D-1}(x^i)^2}\to \infty$, behaves as
\begin{eqnarray}
ds^2&\simeq& \left(-1+\frac{16\pi GM}{(D-2)\Omega_{D-2} r^{D-3}}\right)dt^2-\sum_{i,j=1}^{D-1}\frac{8\pi G J_{ij}}{\Omega_{D-2} r^{D-1}} dt(x^idx^j-x^jdx^i)\nonumber\\
 &&+\left(1+\frac{16\pi GM}{(D-2)(D-3)\Omega_{D-2} r^{D-3}}\right)\sum_{i=1}^{D-1} (dx^i)^2,
\end{eqnarray} 
where $\Omega_{D-2}=2\pi^{(D-1)/2}/\Gamma((D-1)/2)$ is the area of a $(D-2)$-sphere with unit radius,  
 $M$ are the ADM mass, $J_{ij}(=-J_{ji})$ is a angular momentum in a rotational plane $(x^i,x^j)$ and in the choice of suitable coordinates, it can be written in a block diagonal form
\begin{eqnarray}
(J_{ij})=\left( 
\begin{array}{cccccc}
0      &J_{12} &0 &0&0&\cdots \ \\
-J_{12}&0      &0 &0&0&\cdots \\
0&0 &0&J_{34}&0&\cdots \\
0& 0&-J_{34}&0&0&\cdots \\
\vdots&\vdots & \vdots&\vdots&\ddots\\
& & \\
\end{array} 
\right).
\end{eqnarray} 
Following Ref.~\citen{ER-review}, we denote the angular momenta by $J_i:=J_{2i-1,2i}$.

\subsection{Schwarzschild-Tangherlini black hole solutions}\label{sec:Sch}
The generalization of the Schwarzschild solution to higher dimensions was made by Tangherlini in 1963 in a straightforward way~\cite{Tangherlini}. 
The metric is given by
\begin{eqnarray}
ds^2=-\left(1-\frac{\mu}{r^{D-3}}  \right)dt^2+\frac{dr^2}{\left(1-\frac{\mu}{r^{D-3}}  \right)}+r^2d\Omega_{D-2}^2,
\end{eqnarray}
with the metric of a $(D-2)$-dimensional unit sphere
\begin{eqnarray}
d\Omega_{D-2}^2=d\theta_1^2+\sin^2\theta_1d\theta_2^2+\cdots+(\sin^2\theta_1\cdots \sin^2\theta_{D-3})d\theta_{D-2}^2.
\end{eqnarray}
The above metric provides a vacuum solution in the $D$-dimensional Einstein theory, describing static, asymptotically flat higher dimensional black holes.
As seen easily, the black hole horizon exists at $r=r_{h}:=\mu^{1/(D-3)}$. The mass 
is
\begin{eqnarray}
M=\frac{(D-2) \Omega_{D-2} }{16\pi G}\mu. 
%,\quad J_{ij}=0.
\end{eqnarray}

\subsection{Myers-Perry black hole solutions}\label{sec:MP}
The Kerr solution in four-dimensional Einstein theory, which describes asymptotically flat rotating black holes, was generalized to higher-dimensions $(D\ge 5)$ by Myers and Perry~\cite{Myers:1986un} in 1986, who used the Kerr-Schild formalism.
The essential difference from the four dimensional Kerr solution is that there are several independent rotation planes, whose number $N$ depends on the space-time dimensions as $N=[(D-1)/2]$, in higher dimensions. Therefore, 
this solution is specified by the mass parameter $\mu$ and $N$ spin parameters $a_i\ (i=1,\cdots,N)$.  The metric forms in odd and even dimensions are slightly different.

\medskip
When the space-time dimension, $D$, is odd, the metric takes the form
\begin{eqnarray}
ds^2=-dt^2+(r^2+a_i^2)(d\mu_i^2+\mu_i^2d\phi_i^2)+\frac{\mu r^2}{\Pi F}(dt-a_i\mu_i^2d\phi_i)^2+\frac{\Pi F}{\Pi-\mu r^2}dr^2,
\end{eqnarray}
where the functions $F$ and $\Pi$ are defined as 
\begin{eqnarray}
 F(r,\mu_i)=1-\frac{a_i^2\mu_i^2}{r^2+a_i^2},\quad\Pi=\prod_{i=1}^N(r^2+a_i^2).
\end{eqnarray}
and $i=1,\cdots, N$ and $\mu_i$ have to satisfy the constraint $\sum_i\mu_i^2=1$. 
On the other hand, when $D$ is even, the metric is written as 
\begin{eqnarray}
ds^2=-dt^2+(r^2+a_i^2)(d\mu_i^2+\mu_i^2d\phi_i^2)+\frac{\mu r^2}{\Pi F}(dt-a_i\mu_i^2d\phi_i)^2+\frac{\Pi F}{\Pi-\mu r}dr^2,
\end{eqnarray}
where $\mu_i$ satisfies $\sum_i\mu_i^2+\alpha^2=1$ with the constant satisfying $-1\le \alpha \le 1$. The ADM mass and angular momenta for the $i$-th rotational plane $(x_{2i-1},x_{2i})$ are given by
\begin{eqnarray}
M=\frac{(D-2)\Omega_{D-2}}{16\pi G}\mu,\quad J_i
%=J_{2i-1 2i}
=\frac{\Omega_{D-2}}{16 \pi G}\mu a_i.%=\frac{2}{D-2}Ma_i.
\end{eqnarray}

\medskip
It turns out that the horizons exist at the place where $g^{rr}$ vanishes, as in the four-dimensional case. 
For even $D$ with $D\ge 6$, the location of the horizons is determined by the roots of the equation
\begin{eqnarray}
&&\Pi -\mu r=0.\label{eq:even-root}
\end{eqnarray}
The left-hand side is a polynomial of $2N$ order, so that this equation for arbitrary dimensions would have no general analytic solution.
However, a simple consideration enables us to know, at least, the conditions for the existence of the horizon, in particular,  when all spin parameters $a_i$ are non-zero. 
Since curvature singularities exist at $r=0$, the existence of horizons requires that Eq. (\ref{eq:even-root}) should have at least one solution for positive $r$. 
As seen easily, for positive $r$ the function $\Pi-\mu r$ has only a single local minimum, the value of the mass parameter $\mu$ are assumed to be positive.
Now suppose the value of $r$ to be $r_*$ and then it is evident that 
\begin{eqnarray}
\Pi(r_*)-\mu r_* 
\left\{
  \begin{array}{lll}
  >0&\hspace{1cm}\rm{no \ horizon}, \\
  =0&\hspace{1cm}\rm{one\ degenerate\ horizon},  \\ 
  <0&\hspace{1cm}\rm{two\ horizons}.
  \end{array} 
  \right.
\end{eqnarray}

\medskip
For odd $D$ with $D\ge 5$, the horizons exist at the values of $r$ of the equation 
\begin{eqnarray}
&&\Pi -\mu r^2=0.
\end{eqnarray}
The left-hand side is a polynomial of $(D-1)/2$ order in $l:=2r^2$, so that in general for $D=5,7,9$ only, the above equation would have analytic solutions.
In particular, for $D=5$, an analytic solution can be found as
\begin{eqnarray}
r_{\pm}^2=\frac{\mu-a_1^2-a_2^2\pm \sqrt{(\mu-a_1^2-a_2^2)^2-4a_1^2a_2^2}}{2}.
\end{eqnarray}
Therefore  the presence (absence) of the horizon requires the parameters should lie in the range
\begin{eqnarray}
\mu> 0,\quad |a_1|+|a_2| 
\left\{
  \begin{array}{lll}
  >\sqrt{\mu}&\hspace{1cm}\rm{no \ horizon}, \\
  =\sqrt{\mu},\ a_1a_2 =0&\hspace{1cm}\rm{no\ horizon},  \\
  =\sqrt{\mu},\ a_1a_2\not =0&\hspace{1cm}\rm{one\ degenerate\ horizon},  \\ 
  <\sqrt{\mu}&\hspace{1cm}\rm{two\ horizons}.
  \end{array} 
  \right.
\end{eqnarray}
For arbitrary $D$ larger than $5$, when all of spin parameters $a_i$ are non-vanishing, 
the function $\Pi(l)-2\mu l$ has only a single local minimum at $l=l_*$, which is determined from $\partial_{l}\Pi(l_*)-2\mu =0$, and hence from the same discussion, it can be shown that 
\begin{eqnarray}
\Pi(l_*)-2\mu l_* 
\left\{
  \begin{array}{lll}
  >0&\hspace{1cm}\rm{no \ horizon}, \\
  =0&\hspace{1cm}\rm{one\ degenerate\ horizon},  \\ 
  <0&\hspace{1cm}\rm{two\ horizons}.
  \end{array} 
  \right.
\end{eqnarray}
See Ref.~\citen{Myers:1986un} for the more general cases where all of spin parameters are not non-vanishing, in which there is the possibility that the solution has only a single non-degenerate horizon.

\subsection{Stationary and axisymmetric black hole solutions}\label{sec:SA}
In four, or higher dimensions, stationary and axisymmetric solutions of pure gravity were studied.
It was shown by Weyl~\cite{Weyl0} that in four dimensions, static axisymmetric Einstein equation in vacuum can be reduced the Laplace equation in a three-dimensional flat metric and 
further the canonical form of the metric in the four-dimensional stationary asymmetric pure gravity was derived by Papapetrou~\cite{Papapetrou1,Papapetrou2}. 
These works were first generalized to higher dimensions $(D\ge 5)$ by Emparan and Reall~\cite{Weyl}, who showed that in $D$-dimensional pure gravity with $D-2$ commuting Killing vector fields
 the field equation is given in terms of $(D-3)$ solutions of Laplace equations in three-dimensional flat space, and are then generalized to stationary cases by Harmark~\cite{Harmark}, 
who derived a canonical form of the metric and also reduced the Einstein equations to a differential equation on an axisymmetric $(D-2)\times(D-2)$ matrix field in the three-dimensional flat space.

\medskip 
Let us consider a $D$-dimensional spacetime admitting $(D-2)$ commuting Killing vector fields $V_{(i)}\ (i=0,\cdots,D-3)$.
The commutativity of Killing vector fields, $[V_{(i)},V_{(j)}]=0$, enables 
us to find coordinate system $x_i$ $(i=0,\cdots,D-3)$, so that 
$V_{(i)}=\partial/\partial {x^i}$ and the coordinate components of the metric, 
become independent of $x^i$. 
The condition was given in Refs.~\citen{Weyl,Harmark} that the two-dimensional distribution orthogonal 
to $(D-2)$-Killing vector fields $V_{(i)}\ (i=0,\cdots,D-3)$ becomes integrable. 
 We now recall the following generalized Frobenius theorem on the integrability of two-planes orthogonal to Killing vector fields~\cite{Weyl,Harmark}: 

\medskip 
\noindent 
{\bf Theorem.}
{\em Let $V_{(i)}$ $(i=0,\cdots,D-3)$ commuting Killing vector fields such that 
\begin{enumerate}
\item
$V_{(0)}^{[M _0}V_{(1)}^{M _1}\cdots V_{(D-3)}^{M _{D-3}} \nabla^A V_{(i)}^{B ]}=0$
holds at at least one point of the spacetime
for a given $i=0,\cdots,D-3$, 

\item The tensor 
$V_{(i)}^A R_A{} ^{[B}
V_{(0)}^{M _0}V_{(1)}^{M _1}
\cdots V_{(D-3)}^{M_{{\tiny D-3}}]}=0 $ holds for all
$i=0,\cdots ,D-3$,  
\end{enumerate} 
then the two-planes orthogonal to the Killing vector fields 
$V_{(i)}\ (i=0,\cdots,D-3)$ are integrable.}

\medskip 
We are now concerned with a stationary axisymmetric black hole spacetime and hence set the Killing vectors $V_{(0)}=\partial/\partial t$ and $V_{(i)}=\partial/\partial\phi_i\ (i=1,\cdots,D-3)$ to be the asymptotic time translation
Killing vector fields and the rotational Killing vectors with closed integral curves, respectively. The condition 2 holds for arbitrary Ricci-flat spacetime with $D-2$ commuting Killing vector fields.
Furthermore, the axial symmetry of at least one of $V_{(i)} (i=1,\cdots,D-3)$ implies that the condition 1 also holds on the axis of
rotation (fixed points of rotation). Therefore, for any black hole spacetimes with a rotational axis, the two-dimensional surface orthogonal to all the commuting Killing vectors turns out be integrable.
A $D$-dimensional asymptotically flat black hole space-time has at most only $N=[(D-1)/2]$ commuting spacelike Killing vector fields corresponding $U(1)^N$ symmetry.
Therefore, this theorem cannot be applied to asymptotically flat black hole spacetimes with more than five dimensions and hence in the following subsections, we will restrict ourselves to five-dimensional asymptotically flat solutions.

\medskip
It has been shown that for any Ricci-flat
space-time with $D-2$ commuting Killing vector fields $V_{(i)}=\partial/\partial x^i$ from condition of Theorem,
 one can find a coordinate systems such that the metric takes the canonical form
\begin{eqnarray}
ds^2=g_{ij}dx^idx^j+e^{2\nu}(d\rho^2+dz^2),
\end{eqnarray}
with
\begin{eqnarray}
\rho=\sqrt{|{\rm det} g_{ij}|},\label{eq:det=rho}
\end{eqnarray}
where the metric components $g_{ij}=g_{ij}(\rho,z)$ and $\nu=\nu(\rho,z)$ depend on only $\rho$ and $z$.
In terms of the {\it canonical coordinates}, the vacuum Einstein equation $R_{ij}=0$ is written as
\begin{eqnarray}
U_{,\rho}+V_{,z}=0,\label{eq:soliton}
\end{eqnarray}
\begin{eqnarray}
\nu_{,\rho}&=&-\frac{1}{2\rho}+\frac{1}{8\rho}{\rm Tr}\left(U^2-V^2\right),\label{eq:nurho}\\
\nu_{,z}&=&\frac{1}{4\rho}{\rm Tr}\left( UV\right),\label{eq:nuz}
\end{eqnarray}
where the $(D-2)\times(D-2)$ matrices $U$ and $V$ are defined in terms of the $(D-2)$-metric $g=(g_{ij})$  by
\begin{eqnarray}
U=\rho g_{,\rho}g^{-1},\quad V=\rho g_{,z}g^{-1}.
\end{eqnarray}
The integrability of $\nu_{,\rho z}=\nu_{,z\rho}$ is assured by Eq. (\ref{eq:soliton}). For the special case when all the $D-2$ Killing vector fields are orthogonal to each other (when the $(D-2)$-metric has a diagonal form), the canonical form of
the metric is reduced to the generalized Weyl solutions, studied earlier by Emparan and Reall~\cite{Weyl},
\begin{eqnarray}
ds^2=-e^{2U_0}dt^2+\sum_{i=1}^{D-3}e^{2U_i}(dx^i)^2+e^{2\nu}(d\rho^2+dz^2),
\end{eqnarray}
\begin{eqnarray}
\sum_{i=0}^{D-3} U_i=\ln \rho,
\end{eqnarray}
where the $(D-2)$ functions $U_i(\rho,z)\ (i=0,\cdots,D-3)$ are axisymmetric solutions of a Laplace equation in an abstract three-dimensional flat space (namely, $ds^2=d\rho^2+dz^2+\rho^2d\varphi^2$), which is written in the form:
\begin{eqnarray}
\frac{\partial^2 U_i}{\partial \rho^2}+\frac{1}{\rho}\frac{\partial U_i}{\partial \rho}+\frac{\partial^2 U_i}{\partial z^2}=0.
\end{eqnarray}
Eqs.(\ref{eq:nurho}) and (\ref{eq:nuz}) are written as
\begin{eqnarray}
&&\nu_{,\rho}=-\frac{1}{2\rho}+\frac{\rho}{2}\sum_{i-0}^{D-3}\left[( U_{i,\rho})^2-(U_{i,z})^2 \right],\\
&&\nu_{,z}=\rho\sum_{i=0}^{D-3}U_{i,\rho}U_{i,z}.
\end{eqnarray}

\medskip
From the condition (\ref{eq:det=rho}), it immediately turns out that a $(D-2)$-metric $g_{ij}$ has at least one zero eigenvalue on the $z$-axis ($\rho=0$).
As shown in Ref.~\citen{Harmark}, for regular solutions, (by which we here mean that naked curvature singularities do not exist), the matrix $g_{ij}(0,z)$ do not have more than one zero eigenvalue except at isolated points.
Let $a_{k}\ (k=1,\cdots,n)$ ($a_1<a_2<\cdots<a_n$) be the isolated points, which then divide the z-axis into the $n+1$ intervals $[-\infty,a_1],[a_1,a_2],\cdots,[a_n,\infty]$.
The line intervals $[a_{k-1},a_k]\ (k=1,\cdots,n+2,\ a_0=-\infty,a_{n+1}=\infty)$ are called {\it rods} of the solution. In general, a rod $[z_1,z_2]$ such that both of $z_1$ and $z_2$ are finite is called {\it finite rod}, a rod $[z_1,z_2]$ with either $z_1=-\infty$ or $z_2=\infty$ is called  {\it semi-infinite rod}, and further an infinite interval 
$[-\infty,\infty]$ is said to be {\it infinite rod}. Let  $v_{(k)}$ be an eigenvector (so-called rod vector) associated with a zero eigenvalue for a rod $[a_{k-1},a_k]$, namely,
\begin{eqnarray}
g_{ij}(0,z)v_{(k)}^j=0\quad{\rm for}\ z\in[a_{k-1},a_k].
\end{eqnarray} 
When the signature of $\lim _{\rho\to0}g_{ij}(\rho,z)v^iv^j/\rho^2$ is positive, or negative, the rod is said to be {\it timelike}, or {\it spacelike}. In general, a timelike rod corresponds to a horizon (points where a certain linear combination of the Killing vectors $\partial/\partial t+\sum_i \Omega_i \partial/\partial \phi_i$ becomes null, where the constants $\Omega_i\ (i=1,\dots,D-3)$ correspond to the angular velocity of a horizon along $\partial/\partial\phi_i$) and a spacelike rod corresponds to a rotational axis 
( fixed points of an action of a rotational Killing vector). Such a set of rods assigned a corresponding rod vector, $\{[a_{k-1},a_k],\ v_{(k)}\ |\ k=1,\cdots,n+1\}$ is said to be {\it rod structure} of solutions~\cite{Weyl,Harmark}. 

\medskip
Considering the rod structure helps us understand the properties of the solutions such as the global structure or the horizon topology, in particular, it was shown by Hollands and Yazadjive~\cite{Hollands} 
that under symmetry assumptions ${\Bbb R}\times U(1)\times U(1)$, a five-dimensional asymptotically flat black hole spacetime is uniquely determined by the asymptotic conserved charges and rod structure [See Ref. \citen{Hollands} for the precise statement.]. 
Therefore, recently, the concept of the rod structure has been used for constructing physically interesting exact solutions of black holes, combined with the solution-generation techniques.

\subsection{Vacuum black hole solutions in five dimensions}\label{sec:5D}

The topology theorems~\cite{Cai,Helfgott,galloway} yield that 
in five-dimensions, cross-sections of the event horizon must be 
topologically either a sphere, a ring, and a lens-space or their connected sums.
Hollands and Yazadjive\cite{Hollands} showed under symmetry assumptions ${\Bbb R}\times U(1)\times U(1)$, the horizon topology is restricted to either a sphere, a ring, or a lens-space.
The first corresponds to the five-dimensional Schwarzschild-Tangherlini~\cite{Tangherlini} solution, or Myers-Perry solution~\cite{Myers:1986un}, and the second corresponds to the black ring solution~\cite{Emparan:2001wn,MishimaIguchi,Fig,Pom}. 
The third is called  {\it black lens}, which have been not yet found as a regular solution. 
Note that black hole solutions with lens space topologies were considered in earlier works (for example, see Refs. \citen{Gauntlett,IKMT} such as Kaluza-Klein black holes but all of these are non-asymptotically flat. 
In this subsection, we review exact solutions of asymptotically flat black holes in five dimensions which belong to a class of stationary, axisymmetric vacuum solutions in the above sense.

\subsubsection{Black rings with $S^1$-rotation }
As is well known, as for the asymptotically flat, static vacuum black hole solutions of higher-dimensional Einstein equations, 
the Schwarzschild-Tangherlini solution \cite{Tangherlini} is the unique solution \cite{shiromizu}, which is common to 
the four-dimensional case~\cite{uniqueness,Israel,Bunting}. However, this uniqueness for black holes no longer holds for the five-dimensional asymptotically flat, stationary spacetime since
Emparan and Reall~\cite{Emparan:2001wn} discovered the rotating black ring solution whose topology is diffeomorphic to $S^1\times S^2$ %, while 
in addition to the rotating black hole solution with $S^3$ horizon topology found by Myers and Perry \cite{Myers:1986un}.
Actually, as shown by Emparan and Reall~\cite{Emparan:2001wn}, there exist three different stationary black hole solutions in five dimensions, a thin black ring, a fat ring and a black hole, 
for the same mass and angular momentum within a certain parameter region.

\medskip
To keep a balance against its self-gravitational attractive force by centrifugal force, the black ring must be rotating along the $S^1$ direction.
The metric of the black ring rotating along the $S^1$ direction (which is labeled by the angular coordinate $\psi$ given below) can be written in terms of several convenient coordinate systems~\cite{Emparan:2001wn,Emparan2,Harmark}. 
In the $C$-metric coordinates, the metric of the Emparan-Reall solution is given by
\begin{eqnarray}
ds^2&=&-\frac{F(y)}{F(x)}\left(  dt-CR\frac{1+y}{F(y)}d\psi\right)^2\nonumber\\
    & &+\frac{R^2F(x)}{(x-y)^2}\left[-\frac{G(y)}{F(y)}d\psi^2+\frac{G(x)}{F(x)}d\phi^2+\frac{dx^2}{G(x)}-\frac{dy^2}{G(y)}  \right],
\end{eqnarray}
where the functions $F$ and $G$ are defined by
\begin{eqnarray}
F(\xi)=1+\lambda \xi,\quad G(\xi)=(1-\xi^2)(1+\nu \xi),
\end{eqnarray}
and the constant $C$ is 
\begin{eqnarray}
C=\sqrt{\lambda(\lambda-\nu)\frac{1+\lambda}{1-\lambda}}.
\end{eqnarray}
The $(x,y)$ coordinates have the range of 
\begin{eqnarray}
-1\le x\le 1,\quad -\infty <y\le -1, 
\end{eqnarray}
and the parameters lie in the range 
\begin{eqnarray}
0<\nu\le \lambda<1.
\end{eqnarray}
This black ring spacetime admits three mutually commuting Killing vectors, stationary
Killing vector field $\partial_t$, and two independent axial Killing vectors $\partial_\psi$, $\partial_\phi$ with closed integral curves.
It turns out that there exists no closed timelike curves in the domain of outer communication.
Three parameters, $R$, $\nu$ and $\lambda$, are not independent, which comes from the requirement for the absence of the conical singularities.  
To avoid conical singularities at the $\psi$-axis ($y=-1$) and the outer $\phi$-axis ($x=-1$), the coordinates $\psi$ and $\phi$ must have periodicity of 
\begin{eqnarray}
\Delta\phi=\Delta\psi=2\pi\frac{\sqrt{1-\lambda}}{1-\nu}.\label{eq:periodic1}
\end{eqnarray}
This condition also assures that the spacetime is asymptotically flat.
However, even though this condition is satisfied, in general, the spacetime still possesses a disc-shaped conical defect at the inner axis $(x=1)$ of the black ring.
Therefore, the absence of conical singularities at the inner axis ($x=1$) should be required, which imposes the angular coordinate $\phi$ on the periodicity of 
\begin{eqnarray}
\Delta\phi=2\pi\frac{\sqrt{1+\lambda}}{1+\nu}. \label{eq:periodic2}
\end{eqnarray} 
Hence, combining (\ref{eq:periodic1}) and (\ref{eq:periodic2}), one finds that regularity requires that the parameters must satisfy
\begin{eqnarray}
\lambda=\frac{2\nu}{1+\nu^2}.
\end{eqnarray}
This can be interpreted as {\it equilibrium (balance) condition} for a black ring, namely, its radius of the ring is dynamically fixed by the balance between the centrifugal and tensional forces.

\medskip
Unlike the rotating black holes, the rotating black ring spacetime has only a single horizon at $y$ satisfying $G(y)=0$, namely $y=-1/\nu$. 
This solution has two non-vanishing charges, the mass and angular momentum which are given by 
\begin{eqnarray}
M=\frac{3\pi R^2}{4G}\frac{\lambda}{1-\nu},\quad J_\psi:=J_1=\frac{\pi R^3}{2G}\frac{\sqrt{\lambda(\lambda-\nu)(1+\lambda)}}{(1-\nu)^2}.
\end{eqnarray}
 The ergosurface is at $y=-1/\lambda$ where $F(y)=0$.

\subsubsection{Black rings with $S^2$ rotation}
The black ring solution with $S^2$ rotation was first found by Mishima and Iguchi~\cite{MishimaIguchi} by the solitonic method and thereafter
derived by Figueras~\cite{Fig} independently  by the different approach. 
From the discussion of the uniqueness~\cite{Hollands}\footnote{Note that the existence of conical singularities does not affect the proof of the uniqueness as boundary value problem. See also Ref.~\citen{MTY}.}, 
it is now clear that two solutions coincides with each other, but the metric can be written in terms of quite different coordinates.
In the former, the prolate spherical coordinates are used, while in the latter, the $C$-metric coordinates are used. 
Since the latter form is 
simpler 
for expressing a black ring, 
we here follow the work of Figueras (one can also find the expression in terms of the canonical coordinates in Ref.~\citen{TMY}). 
In terms of $C$-metric coordinates, the metric for the $S^2$ rotating black ring is written as
\begin{eqnarray}
ds^2&=&-\frac{H(\lambda,x,y)}{H(\lambda,x,y)}\left(dt-\frac{\lambda ay(1-x^2)}{H(\lambda,y,x)}d\phi \right)^2+R^2\frac{H(\lambda,x,y)}{(x-y)^2}\Bigg[\frac{(1-x^2)F(\lambda,y)}{H(\lambda,y,x)}d\phi^2\cr
&&-\frac{(1-y^2)F(\lambda,x)}{H(\lambda,x,y)}d\psi^2+\frac{dx^2}{(1-x^2)F(\lambda,x)}-\frac{dy^2}{(1-y^2)F(\lambda,y)} \Bigg],
\end{eqnarray}
where the functions, $F$ and $H$, are
\begin{eqnarray}
F(\lambda,\xi)=1+\lambda \xi+\frac{a^2\xi^2}{R^2},\quad H(\lambda,\xi,\eta)=1+\lambda\xi+\frac{a^2\xi^2\eta^2}{R^2},
\end{eqnarray}
and the ranges of $(x,y)$ are the same as those of the Emparan-Reall solution, namely, $-1\le x\le 1$, $-\infty<y\le -1$. 
The outer horizon and inner horizon exist at $y$ satisfying $F(\lambda,y)=0$, 
\begin{eqnarray}
y=y_\pm:=\frac{-\lambda\pm\sqrt{\lambda^2-4a^2/R^2}}{2a^2/R^2}.
\end{eqnarray}
From the requirement for the absence of closed timelike curves
 in the domain of outer communication and the existence of the two horizons, 
the  parameters should satisfy
\begin{eqnarray}
\frac{2a}{R}<\lambda<1+\frac{a^2}{R^2}.
\end{eqnarray}
The condition that there is no conical singularity at the $\psi$-axis ( $y=-1$) and outer $\phi$-axis ($x=-1$) turns to be
\begin{eqnarray}
\Delta\phi=\Delta\psi=\frac{2\pi}{\sqrt{1-\lambda+a^2/R^2}}.\label{eq:condi-1}
\end{eqnarray} 
Furthermore, this condition at the inner axis ($x=1$) is
\begin{eqnarray}
\Delta\phi=\frac{2\pi}{\sqrt{1+\lambda+a^2/R^2}}.\label{eq:condi-2}
\end{eqnarray} 
Note that these two conditions (\ref{eq:condi-1}) and (\ref{eq:condi-2}) cannot be satisfied at the same time, which
means that the existence of canonical singularity cannot be avoided and as a result if one assumes asymptotic flatness,
there necessarily exist conical singularities at the inner axis ($x=1$) to support the horizon.

\medskip
The mass and non-vanishing angular momentum are given by
\begin{eqnarray}
M=\frac{3\pi R^2}{4G}\frac{\lambda}{1-\lambda+a^2/R^2},\quad J_\phi:=J_{2}=-\frac{\pi R^2}{G}\frac{\lambda a}{(1-\lambda+a^2/R^2)^{3/2}}.
\end{eqnarray}

\subsubsection{Black rings with two independent angular momenta}
The five-dimensional vacuum solution which describes {\it balanced} doubly rotating black rings with two independent angular momenta was found by Pomeransky and Sen'kov~\cite{Pom}. 
This solution can be obtained from the more general {\it unbalanced} black ring solution by imposing conical free condition (since the expression of the metric is much more complicated and lengthy, we do not write it here.  Readers can find it in Ref.~\citen{MTY}).
In terms of the $(x,y)$ coordinates,  the metric of the Pomeransky-Sen'kov black ring solution is written
\begin{eqnarray}
ds^2&=&-\frac{H(y,x)}{H(x,y)}(dt+\Omega)^2-\frac{F(x,y)}{H(y,x)}d\psi^2-2\frac{J(x,y)}{H(y,x)}d\phi d\psi+\frac{F(y,x)}{H(y,x)}d\phi^2\nonumber\\
    & &+\frac{2k^2H(x,y)}{(x-y)^2(1-\nu)^2}\left(\frac{dx^2}{G(x)}-\frac{dy^2}{G(y)}\right),
\label{eq:metric}
\end{eqnarray}
where the $1$-form $\Omega$ is given by
\begin{eqnarray}
\Omega&=&-\frac{2k\lambda \sqrt{(1+\nu)^2-\lambda^2}}{H(y,x)}\biggl[\sqrt{\nu}(1-x^2)yd\phi\nonumber\\
& &+\frac{(1+y)d\psi }{(1-\lambda +\nu)(1+\lambda-\nu+\nu(1-\lambda-\nu) x^2y+2\nu x(1-y)} \biggr],
\end{eqnarray}
and, the functions, $G(x),\ H(x,y),\ J(x,y),\ F(x,y)$, are defined by
\begin{eqnarray}
G(x)&=&(1-x^2)(1+\lambda x+\nu x^2),\\
H(x,y)&=&1+\lambda^2-\nu^2+2\lambda \nu(1-x^2)y+2 \lambda x (1-\nu^2y^2)\nonumber\\
      &&+\nu(1-\lambda^2-\nu^2) x^2 y^2,\\
J(x,y)&=& \frac{2k^2\lambda \sqrt{\nu} (1-x^2)(1-y^2)}{(x-y)(1-\nu)^2}(1+\lambda^2-\nu^2+2\lambda \nu(x+y)\nonumber\\
         &&-\nu xy(1-\lambda^2-\nu^2)),\\
F(x,y)&=&\frac{2k^2}{(1-\nu)^2(x-y)^2}\biggl[  G(x)(1-y^2)\{((1-\nu)^2-\lambda^2)(1+\nu)\nonumber\\
        &&+\lambda (1-\lambda^2+2\nu-3\nu^2)y\}+G(y)\{2\lambda^2+\lambda ((1-\nu)^2+\lambda^2)x\nonumber\\
        && +((1-\nu)^2+\lambda^2)(1+\nu)x^2+\lambda(1-\lambda^2-3\nu^2+2\nu^3)x^3\nonumber\\
      &&-\nu(1-\nu)(\lambda^2+\nu^2-1)x^4 \}\biggr]. 
\end{eqnarray}
The two coordinates, $x,y$, run the ranges of $-1\le x\le 1$ and $-\infty<y\le -1$, respectively. 
The solution has three independent parameters satisfying the 
inequalities 
\begin{eqnarray}
0\le \nu<1, \ 2\sqrt{\nu}\le \lambda<1+\nu, \ k>0.
\end{eqnarray}

The horizons, an inner horizon and an outer horizon, exist at the roots of the equation $G(y)=0$, i.e.,
\begin{eqnarray}
\nu y^2+\lambda y +1=0. \label{eq:root}
\end{eqnarray}
 As seen from (\ref{eq:root}), when $\lambda\to 2\sqrt{\nu}$, the outer horizon and inner horizon degenerate and hence this corresponds to the extremal limit.
 The other limit $\nu \to 1$ and $\lambda \to 2$ corresponds to the extremal Myers-Perry solution.

\begin{figure}[!h]
  \begin{center}
   \includegraphics[width=0.45\linewidth]{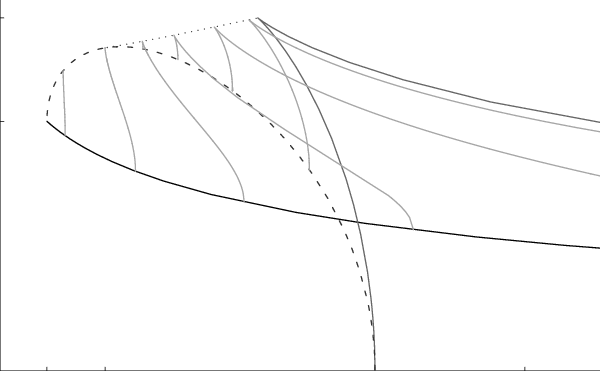}
 \begin{minipage}{0.8\hsize}
  \caption{\small Phase diagrams for the doubly spinning black ring presented in Ref.~\citen{Elvang-R}: 
The horizon area $A_{H}$ (vertical axis) vs.~the angular momentum $j_\psi^2$ (horizontal axis). 
The dark gray curve corresponds to the phase of the Emparan-Reall black ring with $j_\phi=0$, 
which has a cusp where for the fixed $j_\phi$, $j_\psi$ is minimized and the area is maximized and therefore 
there are two branches, the thin ring and the fat ring for the same mass $M$ and angular momentum $J_\psi$.
  Each light gray curve shows branches of constant $j_\phi\not= 0$. 
As $j_\phi$ is increased, the cusp moves to the left, and disappears at $j_\phi=1/5$ and hence there exist only thin ring. 
Each $j_\phi=$constant curve ends at the solid curve and dashed curve.
The solid curve and dashed curve  shows the extremal black ring solution which can be obtained by the limit $\lambda \to 2\sqrt{\nu}$ and the extremal Myers-Perry solution, respectively.  
 }\label{fig:PS-phase}
 \end{minipage}
  \end{center}
\end{figure}

\medskip
The ADM mass and two angular momenta are given by
\begin{eqnarray}
&&M=\frac{3\pi k^2}{G}\frac{\lambda}{1+\nu-\lambda},\quad J_\phi=\frac{4\pi k^3}{G}\frac{\lambda\sqrt{\nu ((1+\nu)^2-\lambda^2)}}{(1+\nu-\lambda)(1-\nu)^2},\\
&&J_\psi=\frac{2\pi k^3}{G}\frac{\lambda(1+\lambda-6\nu+\nu\lambda+\nu^2)\sqrt{(1+\nu)^2-\lambda^2}}{(1+\nu-\lambda)^2(1-\nu)^2},
\end{eqnarray}
which are bounded as
\begin{eqnarray}
j^2_\phi:=\frac{27\pi J_\phi^2}{32G M^3}\le \frac{1}{16},\quad j^2_\psi:=\frac{27\pi J_\psi^2}{32G M^3}\ge \frac{9}{16}.
\end{eqnarray}
As discussed in Ref. \citen{Elvang-R}, (see Fig.~\ref{fig:PS-phase}),  
when $j_\phi^2< 1/25$, the solution has two branches for fixed $j_\phi$, the thin ring branch and the fat ring branch, 
while when $1/25\le j_\phi^2<1/16$, only the thin ring exists in contrast to the Emparan-Reall black ring~\cite{Emparan:2001wn}.

\if0
\begin{figure}[!h]
  \begin{center}
   \includegraphics[width=0.45\linewidth]{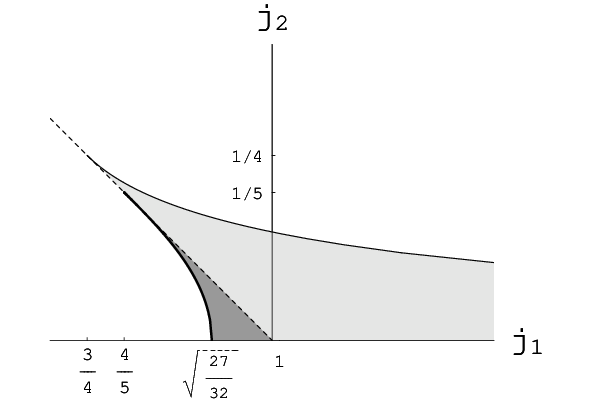}
 \begin{minipage}{0.8\hsize}
  \caption{Phase space of doubly-spinning black rings ($j_1\equiv
j_\psi$, $j_2\equiv j_\phi$) presented in Ref.~\citen{ER-review}.
The dashed line $j_1+j_2=1$ corresponds to the extremal MP
black holes.
The upper thin black curve corresponds to extremal black rings $\lambda \to 2\sqrt{\nu}$.
 The lower thick black curve corresponds to regular non-extremal black rings with minimal spin
$j_1$ along $S^1$ for given $j_2$ on $S^2$. 
Black rings exist in the gray-shaded parameter
regions.
 In the light-gray region there exist only thin
black rings. 
In the dark-gray region, there exist thin
and fat black rings, and MP black holes:
}\label{fig:PS-phase2}
 \end{minipage}
  \end{center}
\end{figure}
\fi

\subsubsection{Black lenses}

Evslin~\cite{Evslin} first tried to construct an asymptotically flat static black lens solution, and however, found that
there necessarily exist naked curvature singularities surrounding
each of the two junctions where the spacelike
rods meet. Moreover, he guessed that these singularities may be resolved by making the black lens rotate. 
Using the inverse scattering method, Chen and Teo~\cite{CT} tried to construct a rotating black lens solution with a single angular momentum, as a result, however, it turned that there must be either conical singularities or naked curvature singularities in the space-time. 
\medskip

The metric of the rotating black lens solution~\cite{CT} is given by
\begin{eqnarray}
ds^2&=&-\frac{H(y,x)}{H(x,y)}(dt+\omega)^2-\frac{F(x,y)}{H(y,x)}d\phi^2-2\frac{J(x,y)}{H(y,x)}d\phi d\psi+\frac{F(y,x)}{H(y,x)}d\psi^2\nonumber\\
    & &+\frac{k^2H(x,y)}{2(1-a^2)(1-b)^3(x-y)^2}\left(\frac{dx^2}{G(x)}-\frac{dy^2}{G(y)}\right),
\label{eq:bl}
\end{eqnarray}
where the $1$-form $\omega$ is
\begin{eqnarray}
\omega&=&\frac{2\kappa}{H(y,x)}\sqrt{\frac{2b(1+b)(b-c)}{(1-a^2)(1-b)}}(1-c)(1+y)\\
&\times&\biggl[\left\{2( 1-b-a^2(1+bx) )^2(1-c)-a^2(1-a^2)b(1-b)(1-x)(1+cx)(1+y)   \right\}d\psi\nonumber\\
  &+&a(1+x)\left\{ a^4(1-c)(1+b)(b-c)+a^2(1-b)(-b+cb+2c)-(1-b)^2c \right\}d\phi \biggr],\nonumber
\end{eqnarray}
and the functions, $G, H, F$ and $J$, are
\begin{eqnarray}
G(x)&=&(1-x^2)(1+cx)\,,\cr
H(x,y)&=&4(1-b)(1-c)(1+bx)\{(1-b)(1-c)-a^2[(1+bx)(1+cy)\cr
&&+(b-c)(1+y)]\}+a^2(b-c)(1+x)(1+y)\cr
&&\times[(1+b)(1+y)\{(1-a^2)(1-b)c(1+x)+2a^2b(1-c)\}\cr
&&-2b(1-b)(1-c)(1-x)]\,,\cr
F(x,y)&=&\frac{2\kappa^2}{(1-a^2)(x-y)^2}\,\Big[4(1-c)^2(1+bx)\{1-b-a^2(1+bx)\}^2G(y)\cr
&&~-a^2G(x)(1+y)^2\Big[\{1-b-a^2(1+b)\}^2(1-c)^2(1+by)\cr
&&-(1-a^2)(1-b^2)(1+cy)\{(1-a^2)(b-c)(1+y)\cr
&&+\{1-3b-a^2(1+b)\}(1-c)\}\Big]\Big]\,,\cr
J(x,y)&=&\frac{4\kappa^2a(1-c)(1+x)(1+y)}{(1-a^2)(x-y)}\,\{1-b-a^2(1+b)\}\{(1-b)c+a^2(b-c)\}\cr
&&~\times\{(1+bx)(1+cy)+(1+cx)(1+by)+(b-c)(1-xy)\}\,.  
\end{eqnarray}
The coordinates $x,y$ take the range $-1\le x\le 1$, $-1/c\le y\le -1$.
The parameters satisfy $0<c\leq b<1$ and $-1<a<1$ and 
%\subsection{Black strings}
to obtain a black lens with horizon topology $L(n,1)$, they must satisfy
\begin{eqnarray}
\frac{2a[(1-b)c+a^2(b-c)]}{[1-b-a^2(1+b)](1-c)}=n\,, \label{eq:lens-condition}
\end{eqnarray}
with a natural number $n$. 
The rod of this solution can be decomposed into four parts: 
(i) $x=-1$:  the outer $\phi$-axis with the rod vector $v=(0,1,0)$, 
(ii) $y=-1/c$:  the event horizon, 
(iii) $x=1$:  the inner $\phi$-axis with the rod vector $v=(0,1,n)$, 
(iv) $y=-1$:  the $\psi$-axis with the rod vector $v=(0,0,1)$. 
From (i) and (ii), one can see that the topology of the horizon is the lens space $L(n;1)$.
The condition for the absence of conical singularities at $x=1$ is
\begin{eqnarray}
m^2\equiv\frac{(1-a^2)^2(1-b)^3(1+c)^2}{[1-b-a^2(1+b)]^2(1+b)(1-c)^2}=1\,.\label{eq:conical-free}
\end{eqnarray}   
The region which satisfies $n>0$ can be decomposed into two regions, $-1<a<-\sqrt{(1-b)/(1+b)}$ and $0<a<\sqrt{(1-b)/(1+b)}$. In the former region, the condition (\ref{eq:conical-free})
 can be satisfied only when $n=2$ but naked curvature singularities always appear around the point $(x,y)=(1,-1)$. 
In the latter region,  this condition cannot be satisfied but there do not exist any naked curvature singularities.
This is why this solution has either naked curvature singularities in the exterior region of the horizon, or conical singularities at the inner axis. However, this does not immediately means that there exists no black lens solution. Note that there may be a black lens solution with a rod structure different from the above solution, or there may exist a less symmetric black lens solution with only ${\Bbb R}\times U(1)$, since the inverse scattering method they used to construct the above solution requires that a spacetime must have ${\Bbb R}\times U(1)\times U(1)$ symmetry.

\subsection{Multiple-black hole solutions}
In five dimensions, multiple-black hole solutions exist as asymptotically flat, stationary,  regular, vacuum solutions, which makes contrast to the four-dimensional case since in four dimensions multi-Kerr black hole solution is well known but conical singularities cannot be avoided between each black holes~\cite{KN,Hoen,DH,eact}. In fact, at least, as for a double black hole system, it has been shown that any configurations of multi-black hole in equilibrium are not admitted by considering boundary value analysis for multi-black holes~\cite{NH}.
This strongly suggests that in four dimensions, the repulsion caused by the spin-spin interaction~\cite{Wald} between rotating black holes is not enough to balance each black hole against gravitational attraction. 
However, with the help of the recent development of solitonic method, in five-dimensional Einstein theory, some multiple-black hole solutions called  {\it black saturn},  {\it black di-ring} and  {\it bicycling black ring} have been constructed one after another. 
Here we provide the brief review on these solutions and summarize their novel properties.

\subsubsection{Black saturn} 
The {\it black saturn} solution describes a rotating black hole surrounded by a concentric rotating black ring.
In terms of the canonical coordinates, the metric of the black saturn solution~\cite{Elvang-Fig} is given by
\begin{eqnarray}
  \label{SaturnMetric}
  ds^2 &=&
  -\frac{H_y}{H_x} \bigg\{dt + \left(\frac{\omega_\psi}{H_y}+q\right)d\psi \bigg\}^2 \cr
  &&+ H_x \bigg\{ k^2 \, P \Big( d\rho^2 + dz^2 \Big) 
       + \frac{G_y}{H_y} \, d\psi^2 + \frac{G_x}{H_x}\, d\phi^2 \bigg\} \, ,
\end{eqnarray}
where the functions, $\omega_\psi$, $G_x$, $G_y$, $P$, $H_x$, $H_y$, are
\begin{eqnarray}
  \omega_\psi
  &=&
  2 \frac{
     c_1\, R_1\, \sqrt{M_0 M_1}
    -c_2\, R_2\, \sqrt{M_0 M_2}
    +c_1^2\,c_2\, R_2\, \sqrt{M_1 M_4}
    -c_1\,c_2^2\, R_1\, \sqrt{M_2 M_4}
  }
  {F \sqrt{G_x}} \, ,
\end{eqnarray}
\begin{eqnarray}
  G_x = \frac{\rho^2\mu_4}{\mu_3\, \mu_5},\quad G_y = \frac{\mu_3\, \mu_5}{\mu_4}, \quad P =  {\cal R}_{34}
^2
      {\cal R}_{15}
{\cal R}_{45},
\end{eqnarray}
\begin{eqnarray}
   H_x &=& F^{-1} \, 
   \bigg[ M_0 + c_1^2 \, M_1 + c_2^2\,  M_2+c_1\, c_2\, M_3 + c_1^2 c_2^2\, M_4 \bigg] \, , \\
    H_y &=& F^{-1} \, 
   \frac{\mu_3}{\mu_4}\, 
   \bigg[ M_0 \frac{\mu_1}{\mu_2} 
   - c_1^2 \, M_1 \frac{\rho^2}{\mu_1\,\mu_2} 
   - c_2^2\,  M_2 \frac{\mu_1\,\mu_2}{\rho^2}
   +  c_1\, c_2\, M_3 
   + c_1^2 c_2^2\, M_4 \frac{\mu_2}{\mu_1} \bigg] \, ,
\end{eqnarray}
with the functions
\begin{eqnarray}
  M_0 &=& \mu_2\, \mu_5^2 {\cal D}_{13}^2 {\cal D}_{24}^2
   {\cal R}_{12}^2{\cal R}_{14}^2
   {\cal R}_{23}^2 \, , \\[2mm]
  M_1 &=& \mu_1^2 \, \mu_2 \, \mu_3\, \mu_4 \, \mu_5 \, \rho^2\,
  {\cal D}_{12}^2 {\cal D}_{24}^2{\cal D}_{15}^2
 {\cal R}_{23}^2  \, , \\[2mm]
  M_2 &=& \mu_2 \, \mu_3\, \mu_4 \, \mu_5 \, \rho^2\,
  {\cal D}_{12}^2 {\cal D}_{13}^2
  {\cal R}_{14}^2{\cal R}_{25}^2  \, ,\\[2mm]
  M_3 &=& 2 \mu_1 \mu_2 \, \mu_3\, \mu_4 \, \mu_5 \,
  {\cal D}_{13}{\cal D}_{15} {\cal D}_{24}
  {\cal R}_{11}{\cal R}_{22}
  {\cal R}_{14}{\cal R}_{23}
  {\cal R}_{25} \, ,\\[2mm]
  M_4 &=& \mu_1^2 \, \mu_2\, \mu_3^2 \, \mu_4^2 \,
  {\cal D}_{15}^2
  {\cal R}_{12}^2{\cal R}_{25}^2  \, ,
\end{eqnarray}
and
\begin{eqnarray}
  F &=& \mu_1\, \mu_5\,  {\cal D}_{13}^2{\cal D}_{24}^2
  {\cal R}_{23}
  {\cal R}_{14}
 {\cal R}_{24}
  {\cal R}_{25}
 {\cal R}_{35}
  \prod_{i=1}^5 {\cal R}_{ii} \, .
\end{eqnarray}
Here the functions, $\mu_i$, $R_i$ and ${\cal R}_{ij}$, are defined by
\begin{eqnarray}
\mu_i=\sqrt{\rho^2+(z-a_i)^2}-(z-a_i),
\end{eqnarray}
\begin{eqnarray}
R_i = \sqrt{\rho^2 + (z-a_i)^2},\quad {\cal R}_{ij}=\rho^2+\mu_i\mu_j,\quad{\cal D}_{ij}=\mu_i-\mu_j,
\end{eqnarray}
and $a_i(i=1,\cdots,5),\ c_1,\ c_2,\ q$ and $k$ are constants, satisfying the inequalities $a_1< a_5< a_4 < a_3 < a_2$.

\medskip
The above solution apparently has nine parameters, $(a_i,c_1,c_2,q,k)$  but all of these are not independent.
Following the discussion in Ref.~\citen{Elvang-Fig}, we now summarize the number of physically meaning parameters. 
As well known, in the canonical coordinate system, there is the degrees of freedom in the choice of the coordinate $z$, i.e., degrees of freedom with respect to
shift translation $z \to z + \alpha$, which means that using this can reduce the five parameters, $a_i$, to four. 
To fix this gauge and extract a scale, according to  Ref.~\citen{Elvang-Fig}, let us here define the  dimensionless coordinate $\bar z$ by
\begin{eqnarray}
z=L^2 \bar z+a_1,
\end{eqnarray}
and introduce the dimensionless parameters by
\begin{eqnarray}
\kappa_i=\frac{a_{i+2}-a_1}{L^2}, \quad \bar c_2=\frac{c_2}{c_1(1-\kappa_2)}\ (i=1,2,3).
\end{eqnarray}
where $L^2:=a_2-a_1$ is a overall scale. Here, the constants $k_i$ are ordered as $0< \kappa_3 < \kappa_2<\kappa_1<1$. 

\noindent
(i) {\sl Asymptotic flatness}: The constants $k$ and $q$ are determined so that the space-time satisfies asymptotic flatness as
\begin{eqnarray}
k=\frac{1}{|1+\kappa_2\bar c_2|},\quad q=\sqrt{\frac{2\kappa_1\kappa_2(a_2-a_1)}{\kappa_3}}\frac{\bar c_2 }{1+\kappa_2\bar c_2}.
\end{eqnarray}

\noindent
(ii) {\sl Regularity}: Curvature singularities appears at $(\rho,\bar z)=(0,0)$ (i.e., $(\rho,z)=(0,a_1)$) on the rod, but this can be removed by imposing the following condition on the parameters.
\begin{eqnarray}
|c_1|=L\sqrt{\frac{2\kappa_1\kappa_2}{\kappa_3}}.\label{eq:reg}
\end{eqnarray}
After imposing this condition on the parameters, the rod structure of the above  black saturn solution can be summarized as follows:
\begin{itemize}
\item The semi-infinite spacelike rod $(-\infty,\kappa_3]$ with the rod vector $v=(0,1,0)$, which corresponds to the $\phi$-axis. 
To avoid conical singularities on $\rho=0,\ \bar z\in(-\infty,\kappa_1] $, i.e., to assure asymptotic flatness,  the angular coordinate $\phi$ should the period $\triangle \phi=2\pi$.

\item The finite timelike rod $[\kappa_3, \kappa_2]$ with the rod vector $v=(1,0,\Omega_\psi^{BR})$ corresponds to the horizon of a black ring which 
 has the $S^1$ parameterized by the angular coordinate $\psi$ and the $S^2$ by the coordinates $(z,\phi)$, 
where $\Omega_\psi^{BR}$ denotes the angular velocities along the $S^1$-direction (the $\psi$-direction) of the  rotating black ring and is explicitly written as
\begin{eqnarray}
\Omega^{BR}_\psi
  &=& 
  \frac{1+\kappa_2\,\bar c_2}{L}
  \sqrt{\frac{\kappa_1 \kappa_3}{2 \kappa_2}} 
    \frac{\kappa_3  - \kappa_2 (1-\kappa_3) \bar c_2}
     {\kappa_3 -  \kappa_3 (\kappa_1 -\kappa_2) \bar c_2 + \kappa_1 \kappa_2 (1-\kappa_3) \bar c_2^2}.
\end{eqnarray}
  
\item The finite spacelike rod $[\kappa_2,\kappa_1]$ with the rod vector $(0,1,0)$, which corresponds to the $\phi$-axis between the black hole and the black ring.

\item The finite timelike rod $[\kappa_1, 1]$ with the rod vector $v=(1,0,\Omega_\psi^{BH})$ corresponds to the horizon of a black hole, 
where  $\Omega_\psi^{BH}$ denotes the angular velocity along the $\psi$-direction of the  rotating black hole and are written as
\begin{eqnarray}
\Omega^{BH}_\psi
  &=& 
  \frac{1+\kappa_2\,\bar c_2}{L}
  \sqrt{\frac{\kappa_2 \kappa_3}{2 \kappa_1}} \, 
    \frac{\kappa_3 (1-\kappa_1) - \kappa_1 (1-\kappa_2) (1-\kappa_3) \bar c_2}
     {\kappa_3(1-\kappa_1) + \kappa_1 \kappa_2 (1-\kappa_2) (1-\kappa_3) \bar c_2^2}. 
\end{eqnarray}

\item The semi-infinite spacelike rod $[1,\infty)$ with the rod vector $v=(0,0,1)$. 
To avoid conical singularities on $\rho=0,\ \bar z\in [\kappa_5,\infty)$, i.e., to assure asymptotic flatness,  the angular coordinate $\psi$ should the period $\triangle \psi=2\pi$.

\end{itemize}

\noindent
(iii) {\sl Balance condition}: Even if the condition (\ref{eq:reg}) is satisfied, in general, the solution is still not regular, i.e., conical singularities exist the $\phi$-axis between the black hole and the black ring. 
The existence of conical singularities can be avoid if the parameters satisfy the equation:
\begin{eqnarray}
  \bar c_2 = \frac{1}{\kappa_2}
  \left[
   \epsilon
   \frac{\kappa_1-\kappa_2}
    { \sqrt{\kappa_1 (1-\kappa_2)(1-\kappa_3)(\kappa_1-\kappa_3)} } - 1
  \right] \, , 
\end{eqnarray}
with
\begin{eqnarray}
\left\{
  \begin{array}{ccc}
  \epsilon = +1 & 
  ~\rm{when}~ &\bar c_2 > -\kappa_2^{-1}  \\ 
  \epsilon = -1 &~\rm{when}~ & \bar c_2 < -\kappa_2^{-1}
  \end{array} 
   \right. 
  \, . 
\end{eqnarray}

Thus, as a result, from regularity and the balance condition, the black saturn solution has four independent parameters, which roughly corresponds to physical degrees of freedom of the mass, angular momentum along the $\psi$ direction for the black hole and the mass, angular momentum along the $S^1$ direction for the black ring.

\medskip
We briefly summarize the physical properties of the balanced black saturn solution which is studied in Ref.~\citen{Elvang-Fig}.
\begin{itemize}

\item {\sl Co-rotation and counter-rotation}: The black hole and black ring have independent angular momenta and therefore they can rotate both in the same direction ({\it co-rotation}) and in the opposite directions ({\it counter-rotation}). In particular, when they are counter-rotating, the total ADM  angular momenta measured at infinity can vanish, which means that in five dimensions, the Schwarzschild-Tangherlini solution is not the only asymptotically flat vacuum solution with vanishing angular momenta in to contrast to the four-dimensional case.

\item {\sl Continuous non-uniqueness}: For the same total mass and angular momenta, the balanced black saturn solution  can admit continuous configuration of a black hole and a black ring.  The solution exhibits $2$-fold continuous non-uniqueness. Furthermore, it also exhibits discrete non-uniqueness, i.e., there exists a thin ring branch and a fat ring branch within a certain parameter region.

\item {\sl Frame-dragging}: For the black saturn, there is mutual gravitational interaction between the black hole and black ring and hence one has a physically interesting effect on the other through so-called {\it frame-dragging}. For instance, even if the black hole at the center of the black ring has zero angular momenta, the black hole can be rotating  in the sense that the angular velocity of the black hole is non-vanishing, which can be considered to be the effect of dragging caused by the co-rotating black ring ({\it rotational frame-dragging}). Further, even if the black hole has a non-zero angular momentum, the black hole can be non-rotating, i.e., the angular velocity of the black hole vanishes, by the effect of dragging caused by the counter-rotating black ring ({\it counter frame-dragging}).

\end{itemize}

\subsubsection{Black di-ring}
The black di-ring solution was first discovered by Iguchi and Mishima~\cite{diring} by using B\"acklund transformation, 
and shortly afterward the solution was re-derived by Evslin and Krishnan~\cite{Evslin-Krishnan} by using the inverse scattering method.
The two expressions for the black di-ring solution in Refs. \citen{diring} and \citen{Evslin-Krishnan} seem to take quite different forms even in terms of the same coordinate system. 
However, they were numerically shown to be equivalent in Ref.~\citen{thermo-diring}, which including their parameter ranges that the regular solutions should have.
%As used in the black saturn solution, it may be more convenient to use the canonical coordinates when one considers, for instance,  the rod structure~\cite{Harmark} of a spacetime, and 
We here write down the expression for the metric derived in Ref.~\citen{Evslin-Krishnan}.  
%(even if one performs the coordinate transformation from the expression in Ref.~\citen{diring} into the canonical coordinates used in Ref.~\citen{Evslin-Krishnan},  

\medskip
The metric of the black di-ring solution given in Ref.~\citen{Evslin-Krishnan} is written as
\begin{eqnarray}
\label{ringmetric}
ds^2&=&\frac{X_1+c_1^2X_2+c_2^2X_3+2c_1c_2X_4+c_1^2c_2^2X_5}{\mu_3\mu_6\Delta}dt^2\cr
&&+2\frac{-c_1Y_1-c_2Y_2+c_1^2c_2Y_3+c_1c_2^2Y_4}{\Delta}dtd\psi\\
&&+\frac{Z_1+c_1^2Z_2+c_2^2Z_3+2c_1c_2Z_4+c_1^2c_2^2Z_5}{\mu_1\mu_4\Delta}d\psi^2+\frac{\mu_3\mu_6\rho^2}{\mu_2\mu_5\mu_7}d\phi^2\cr
&&+\frac{(A_1+c_1^2 A_2+2c_1c_2
A_3+c_1^2c_2^2A_4+c_2^2A_5)(d\rho^2+dz^2)}{\mu_1\mu_2\mu_4\mu_5{\cal D}_{14}^2
{\cal R}_{13}{\cal
R}_{14}^2{\cal R}_{16}{\cal
R}_{17}{\cal R}_{25}^2{\cal R}_{27}^2{\cal
R}_{34}{\cal R}_{36}^2{\cal R}_{46}{\cal R}_{47}{\cal
R}_{57}^2\prod_{i=1}^{7}{\cal R}_{ii}},\nonumber
\end{eqnarray}
where the functions, $A_i$, $X_i$, $Y_j\  $, $Z_i$ $(i=1,\cdots,5,\ j=1,\cdots,4)$, are defined by 
\begin{eqnarray}
A_1&=&\mu_2^2 \mu_5^2 \mu_7 {\cal D}_{14}^2 {\cal R}_{12}
{\cal R}_{13}^2{\cal R}_{15}{\cal R}_{16}^2{\cal R}_{17}^2{\cal
R}_{23}{\cal
R}_{24}{\cal R}_{26}{\cal R}_{34}^2{\cal R}_{35}{\cal R}_{37}{\cal
R}_{45}{\cal R}_{46}^2{\cal R}_{47}^2{\cal R}_{56}{\cal R}_{67}, \nonumber
\\
A_2&=&\mu_1^2\mu_2\mu_3\mu_5\mu_6\mu_7^2\rho^2{\cal D}_{12}^2{\cal
D}_{15}^2
{\cal R}_{12}{\cal R}_{14}^2{\cal
R}_{15}{\cal R}_{23}{\cal
R}_{24}{\cal R}_{26}{\cal R}_{34}^2{\cal
R}_{35}{\cal R}_{37}{\cal R}_{45}{\cal R}_{46}^2{\cal
R}_{47}^2{\cal R}_{56}{\cal R}_{67}, \nonumber \\
A_3&=&\mu_1\mu_2\mu_3\mu_4\mu_5\mu_6\mu_7^2 \rho^2{\cal D}_{12}{\cal
D}_{15}{\cal D}_{24}{\cal D}_{45} \nonumber \\
&&\times{\cal R}_{12}{\cal R}_{14}^2{\cal R}_{15}{\cal
R}_{23}{\cal R}_{24}{\cal R}_{26}{\cal R}_{34}^2{\cal R}_{35}{\cal R}_{37}{\cal
R}_{45}{\cal R}_{46}^2{\cal R}_{47}^2{\cal R}_{56}{\cal R}_{67},\nonumber \\
A_4&=&\rho^{8}\mu_1^2\mu_3^2\mu_4^2\mu_5\mu_6^2\mu_7^3{\cal
D}_{12}^2{\cal
D}_{14}^2{\cal D}_{15}^2{\cal D}_{24}^2{\cal
D}_{45}^2{\cal R}_{12}{\cal R}_{15}{\cal
R}_{23}{\cal R}_{24}{\cal R}_{26}{\cal R}_{35}{\cal
R}_{37}{\cal R}_{45}{\cal R}_{56}{\cal R}_{67}, \nonumber\\
A_5&=&\rho^2\mu_2\mu_3\mu_4^2\mu_5\mu_6\mu_7^2{\cal
D}_{24}^2{\cal
D}_{45}^2{\cal
R}_{12}{\cal
R}_{13}^2{\cal
R}_{14}^2{\cal R}_{15}{\cal
R}_{16}^2{\cal R}_{17}^2{\cal R}_{23}{\cal
R}_{24}{\cal R}_{26}{\cal R}_{35}{\cal R}_{37}{\cal R}_{45}{\cal
R}_{56}{\cal R}_{67},
\nonumber
\end{eqnarray}
with the functions $\Delta\equiv D_1+c_1^2
D_2+c_2^2D_3+2c_1c_2D_4+c_1^2c_2^2D_5$, 
\begin{eqnarray}
&&X_1=-\mu_1\mu_4D_1, \ X_2=\rho^2\mu_1^{-1}\mu_4D_2, 
X_3=\rho^2\frac{\mu_1}{\mu_4}D_3, \ X_4=\rho^2D_4, \
X_5=-\frac{\rho^4}{\mu_1\mu_4}D_5,\nonumber
\end{eqnarray}
\begin{eqnarray}
Y_1&=&\mu_2^2\mu_5^2\mu_7{\cal D}_{12}{\cal D}_{14}{\cal D}_{15}{\cal R}_{11}{\cal R}_{13}{\cal R}_{14}{\cal R}_{34}{\cal R}_{16}{\cal R}_{46}^2{\cal R}_{17}{\cal R}_{47}^2,\cr
Y_2&=&\mu_2^2\mu_5^2\mu_7{\cal D}_{14}{\cal D}_{24}{\cal D}_{45}{\cal R}_{44}{\cal R}_{13}^2{\cal R}_{14}{\cal R}_{34}{\cal R}_{16}^2{\cal R}_{46}{\cal R}_{17}^2{\cal R}_{47}, \cr
Y_3&=&\mu_1^2\mu_2\mu_3\mu_5\mu_6\mu_7^2\rho^4{\cal D}_{12}^2{\cal D}_{14}{\cal D}_{24}{\cal D}_{15}^2{\cal D}_{45}{\cal R}_{44}{\cal R}_{14}{\cal R}_{34}{\cal R}_{46}{\cal R}_{47}, \cr
Y_4&=&\mu_2\mu_3\mu_4^2\mu_5\mu_6\mu_7^2\rho^4{\cal D}_{12}{\cal D}_{14}{\cal D}_{24}^2{\cal D}_{15}{\cal D}_{45}^2{\cal R}_{11}{\cal R}_{13}{\cal R}_{14}{\cal R}_{16}{\cal R}_{17},\nonumber
\end{eqnarray}
\begin{eqnarray}
&&Z_1=\mu_2\mu_5\mu_7 D_1, \ Z_2=-\frac{\mu_1^2\mu_2\mu_5\mu_7}{\rho^2}D_2,\ Z_3=-\frac{\mu_2\mu_4^2\mu_5\mu_7}{\rho^2}D_3, \ \cr
&&Z_4=-\frac{\mu_1\mu_2\mu_4\mu_5\mu_7}{\rho^2}D_4, \ Z_5=\frac{\mu_1^2\mu_2\mu_4^2\mu_7}{\mu_5\rho^4}D_5, \nonumber
\end{eqnarray}
\begin{eqnarray}
D_1&=&\mu_2^2\mu_5^2{\cal D}_{14}^2
{\cal R}_{13}^2{\cal R}_{34}^2{\cal R}_{16}^2{\cal
R}_{46}^2{\cal R}_{17}^2{\cal R}_{47}^2, 
\cr
D_2&=&\mu_1^2\mu_2\mu_3\mu_5\mu_6\mu_7\rho^2{\cal D}_{12}^2{\cal D}_{15}^2
{\cal R}_{14}^2{\cal R}_{34}^2{\cal R}_{46}^2{\cal R}_{47}^2,
\cr
D_3&=&\mu_2\mu_3\mu_4^2\mu_5\mu_6\mu_7\rho^2{\cal D}_{24}^2{\cal
D}_{45}^2
{\cal R}_{13}^2{\cal R}_{14}^2{\cal R}_{16}^2{\cal
R}_{17}^2,
\cr
D_4&=&\mu_1\mu_2\mu_3\mu_4\mu_5\mu_6\mu_7\rho^2{\cal D}_{12}{\cal D}_{24}
{\cal D}_{15}{\cal D}_{45}{\cal R}_{11}{\cal
R}_{44}{\cal R}_{13}{\cal R}_{34}{\cal R}_{16}{\cal R}_{46}{\cal
R}_{17}{\cal R}_{47}, 
\cr
D_5&=&\mu_1^2\mu_3^2\mu_4^2\mu_6^2\mu_7^2\rho^8 {\cal D}_{12}^2{\cal
D}_{14}^2{\cal D}_{24}^2{\cal D}_{15}^2
{\cal D}_{45}^2.\nonumber
\end{eqnarray}

\medskip
This black di-ring solution apparently includes the nine parameters ($a_i$, $c_1$, $c_2$).  In general, it is not regular and cannot be in equilibrium. 
As discussed in the black saturn (the more detail discussion can be found in Refs.~\citen{diring,thermo-diring,Evslin-Krishnan}), let us count how many independent parameters the solution has.
As mentioned in the black saturn, to fix the gauge and scale,  let us introduce the  dimensionless coordinate $\bar z$ and the dimensionless parameters $\kappa_i\  (i=1,\cdots,6) $ by
\begin{eqnarray}
z=L^2 \bar z+a_1,
\end{eqnarray}
\begin{eqnarray}
\kappa_i=\frac{a_{i+1}-a_1}{L^2}, 
\end{eqnarray}
where $L^2:=a_7-a_1$ is a overall scale and then they satisfy $0< \kappa_1< \kappa_2<\kappa_3<\kappa_4<\kappa_5<\kappa_6<1$. 

\noindent
(i) {\sl Regularity}: 
 For the above black di-ring solution, in general, there are singularities $\bar z=0$ ($z=a_1$) and $\bar z=\kappa_3$ ($z=a_4$), where the metric components of $g_{tt}$ and $g_{\psi\psi}$ blow up and hence they yield curvature singularities just there. 
To get rid of these, the parameters must be set
\begin{eqnarray}
c_1= L\sqrt{\frac{2\kappa_2\kappa_5\kappa_6}{\kappa_1\kappa_4}}, \quad c_2= L\sqrt{\frac{2(\kappa_3-\kappa_2)(\kappa_5-\kappa_3)(\kappa_6-\kappa_3)}{(\kappa_3-\kappa_1)(\kappa_4-\kappa_3)}}. \label{eq:reg}
\end{eqnarray}

After imposing this condition on the parameters, the rod structure of the above  black saturn solution can be summarized as follows:
\begin{itemize}
\item The semi-infinite spacelike rod $(-\infty,\kappa_1]$ with the rod vector $v=(0,1,0)$, which corresponds to the $\phi$-axis. 
To avoid conical singularities on $\rho=0,\ \bar z\in(-\infty,0] $, the angular coordinate $\phi$ should have the period $\triangle \phi=2\pi$.
By the similar discussion, for the rod $\rho=0,\ \bar z\in(0,\kappa_1] $, the periodic of the angular coordinate $\phi$ should be set
\begin{eqnarray}
\Delta \phi=\frac{2
\pi c_1}{L}\sqrt{\frac{\kappa_1\kappa_4}{2\kappa_2\kappa_5\kappa_6}}.
\end{eqnarray}
From (\ref{eq:reg}), one can see that this yields $2\pi$.

\item The finite timelike rod $[\kappa_1, \kappa_2]$ with the rod vector $v=(1,0,\Omega_\psi^{BR_{out}})$ corresponds to the horizon of a black ring whose
 has the $S^1$ parameterized by the angular coordinate $\psi$ and the $S^2$ by the coordinates $(z,\phi)$, with $\Omega_\psi^{BR_{out}}$ denotes the angular velocity of the outer black ring.

\item The finite spacelike rod $[\kappa_2,\kappa_4]$ with the rod vector $(0,1,0)$, which corresponds to the $\phi$-axis between  two black rings.

\item The finite timelike rod $[\kappa_4, \kappa_5]$ with the rod vector $v=(1,0,\Omega_\psi^{BR_{in}})$ corresponds to the horizon of a black hole, 
 has the $S^1$ parameterized by the angular coordinate $\psi$ and the $S^2$ by the coordinates $(z,\phi)$, with $\Omega_\psi^{BR_{in}}$ denotes the angular velocity of the inner black ring.

\item The semi-infinite spacelike rod $[\kappa_6,\infty)$ with the rod vector $v=(0,0,1)$. 
To avoid conical singularities on $\rho=0,\ \bar z\in [\kappa_5,\infty)$, i.e., to assure asymptotic flatness,  the angular coordinate $\psi$ should the period $\triangle \psi=2\pi$.

\end{itemize}

\noindent
(ii) {\sl Balance condition}: Even under the above conditions (\ref{eq:reg}), the black di-ring cannot be in equilibrium, i.e., 
there are still conical singularities in $\rho=0,\bar z\in[\kappa_2,\kappa_3]$,  $\rho=0,\bar z\in[\kappa_3,\kappa_4]$ between 
the two black rings  and further in $\rho=0,\bar z\in[\kappa_5,\kappa_6]$, on the $\phi$-axis of the inner black ring 
and hence to avoid these singularities, one needs to impose the following periodicities on the coordinate $\phi$, respectively
\begin{eqnarray}
\Delta \phi=2\pi\frac{|Y+c_1c_2Z|}{\sqrt{X}}= 2\pi
\end{eqnarray}
for $\rho=0,\ \bar z\in[\kappa_2,\kappa_3]$,
\begin{eqnarray}
\Delta\phi=2\pi\frac{|c_1U+c_2V|}{\sqrt{W}}=2\pi,
\end{eqnarray}
for $\rho=0,\ \bar z\in[\kappa_3,\kappa_4]$, and
\begin{eqnarray}
\Delta\phi=2\pi\sqrt{\frac{\kappa_6(\kappa_6-\kappa_3)(\kappa_6-\kappa_2)(\kappa_6-\kappa_5)}{(\kappa_6-\kappa_1)^2(\kappa_6-\kappa_4)^2}}=2\pi,
\end{eqnarray}
for $\rho=0,\ \bar z\in[\kappa_5,\kappa_6]$, where
\begin{eqnarray}
X&=&\frac{4\kappa_4^2(\kappa_5-\kappa_2)^2(\kappa_6-\kappa_3)^2(\kappa_7-\kappa_2)^2(\kappa_4-\kappa_3)\kappa_6\kappa_7}{(\kappa_4-\kappa_2)\kappa_5(\kappa_5-\kappa_3)(\kappa_6-\kappa_2)(\kappa_7-\kappa_3)}, \nonumber \\
Y&=&2(\kappa_3-\kappa_2)\kappa_5\kappa_6,  \\
Z&=&\kappa_1(\kappa_4-\kappa_3),\nonumber
\end{eqnarray}
\begin{eqnarray}
U&=&\kappa_1(\kappa_5-\kappa_3)(\kappa_6-\kappa_3), \nonumber \\
V&=&(\kappa_3-\kappa_1)\kappa_5\kappa_6, \\
W&=&\frac{2\kappa_3^2(\kappa_4-\kappa_1)^2\kappa_5(\kappa_5-\kappa_2)^2(\kappa_5-\kappa_3)\kappa_6(\kappa_6-\kappa_1)^2(\kappa_6-\kappa_3)}{\kappa_4(\kappa_4-\kappa_3)(\kappa_4-\kappa_2)(\kappa_5-\kappa_1)(\kappa_6-\kappa_2)}.\nonumber
\end{eqnarray}
These apparently seem to yield three constraints for the parameters but only two are independent, which one can see from the regularity conditions~(\ref{eq:reg})\footnote{There seem to be typos in Ref.~\citen{Evslin-Krishnan} and hence here we have corrected them~\cite{private}.}.

\medskip
Thus to summarize, from regularity and the balance conditions for the di-ring,  the balanced black di-ring solution turns out to have four independent parameters, 
which corresponds to physical degrees of freedom of each mass, each angular momentum along the $S^1$ direction (labeled by $\psi$-direction) for two black rings.

\subsubsection{Bicycling black rings}
The asymptotically flat, stationary vacuum solution which describes two spinning black rings placed in orthogonal independent rotational planes ---what is called {\it bicycling black rings} (or, {\it orthogonal black rings})--- was presented in Refs.~\citen{Izumi,Elvang-R} and therein its detailed analysis was provided.
The bicycling black ring solution can be written as
\begin{eqnarray}
ds^2 &=& - \frac{H_y}{H_x} \left( dt - \frac{\omega_\phi}{H_y} \, d\phi
      - \frac{\omega_\psi}{H_y} \, d\psi  \right)^2\cr
     &&+ H_y^{-1} \left( 
         G_x \, d\phi^2
        +G_y \, d\psi^2
        -2 J_{xy}\, d\phi \, d\psi \right) + P \,  H_x   \left( d\rho^2 + dz^2 \right) \, ,
\end{eqnarray}
\begin{eqnarray}
\omega_\psi &=&
  c_1  \, \frac{ {\cal R}_{11} }{\mu_1}
  \sqrt{\frac{\mu_2 \, \mu_4 \, \mu_6}{\mu_3 \, \mu_7 \, \rho^2}}
  \Big( \sqrt{M_0\, M_1}
  + b_2^2 \, \frac{\mu_7}{\rho} \sqrt{M_2 \, M_3}  \Big)
  \, , \\[2mm]
  \omega_\phi &=&
  b_2  \, \frac{Z_{77}}{\mu_7}
  \sqrt{\frac{\mu_1 \, \mu_5}{\mu_2 \, \mu_4 \, \mu_6}}
  \Big( \sqrt{M_0\, M_2}
  + c_1^2 \, \frac{\rho}{\mu_1} \sqrt{M_1 \, M_3}  \Big) \, ,
\end{eqnarray}
\begin{eqnarray}
P = \frac{\mu_2 \, {\cal R}_{23} \, {\cal R}_{25} \, {\cal R}_{34} \,  {\cal R}_{35} \, {\cal R}_{36}
    \, {\cal R}_{45} \, {\cal R}_{47} \, {\cal R}_{56} \, {\cal R}_{57} \, {\cal R}_{67} }
   {\mu_1\,  \mu_5^4\,  \mu_7 \, {\cal D}_{37}^4
    \, {\cal R}_{12} \, {\cal R}_{13} \, {\cal R}_{14} \, {\cal R}_{15}^2
    \, {\cal R}_{16} \, {\cal R}_{17} \, {\cal R}_{24}^2 \, {\cal R}_{26}^2 \, {\cal R}_{27}
    \, {\cal R}_{37}^2 \, {\cal R}_{46}^2 \,
    \prod_{i=1}^7 {\cal R}_{ii}  \, , } \, ,
\end{eqnarray}
\begin{eqnarray}
H_x &=&
    \Big( M_0 + c_1^2  \, M_1 + b_2^2
      \, M_2 - c_1^2  \, b_2^2 \,  M_3 \Big) \, , \\[3mm]
    H_y &=& \frac{\mu_5}{\mu_3}
    \Big( \frac{\mu_1}{\mu_7} M_0
    - c_1^2  \, \frac{\rho^2}{\mu_1\,\mu_7} M_1
    - b_2^2  \, \frac{\mu_1\,\mu_7}{\rho^2} M_2
    - c_1^2  \, b_2^2 \,   \frac{\mu_7}{\mu_1} M_3 \Big) \, ,\\[3mm]
    G_x &=&
    \frac{\mu_1\, \mu_5 \, \rho^2}{\mu_2\,\mu_4\,\mu_6}
    \Big( M_0
    - c_1^2  \,\, \frac{\rho^2}{\mu_1^2}\, M_1
    + b_2^2  \, M_2
    + c_1^2  \, b_2^2 \,\,  \frac{\rho^2}{\mu_1^2} \, M_3 \Big) \, ,
    \\[3mm]
    G_y &=&
    \frac{\mu_2\,\mu_4\,\mu_6}{\mu_3\, \mu_7}
    \Big( M_0
    + c_1^2  \, M_1
    - b_2^2  \,\, \frac{\mu_7^2}{\rho^2}\, M_2
    + c_1^2  \, b_2^2 \,\, \frac{\mu_7^2}{\rho^2}\, M_3 \Big) \, ,
\end{eqnarray}
\begin{eqnarray}
J_{xy} &=& c_1 \, b_2 \;
     \rho^2 \, \mu_1 \mu_2 \, \mu_3 \, \mu_4 \, \mu_5^2 \, \mu_6 \,
       {\cal D}_{37}^2 {\cal D}_{47}
      {\cal D}_{57} {\cal D}_{67} 
       \, {\cal R}_{11} \, {\cal R}_{77} \, {\cal R}_{12} \, {\cal R}_{13} \, {\cal R}_{14}\cr
       &&\times
         \, {\cal R}_{15}^2 \, {\cal R}_{16} \, {\cal R}_{17} \, {\cal R}_{27} \, ,
\end{eqnarray}
\begin{eqnarray}
M_0 &=&  \mu_4 \, \mu_5^3 \, \mu_6 \, \mu_7 \,  {\cal D}_{37}^4
   \, {\cal R}_{12}^2 \, {\cal R}_{13}^2 \, {\cal R}_{14}^2 \, {\cal R}_{16}^2
    \, {\cal R}_{17}^2 \, {\cal R}_{27}^2  \, , \\[2.5mm]
  M_1 &=& \rho^2\, \mu_1^2 \, \mu_2 \, \mu_3\, \mu_4^2 \,
     \mu_5 \, \mu_6^2 \, {\cal D}_{17}^2 \,  {\cal D}_{37}^4
  \, {\cal R}_{15}^4 \, {\cal R}_{17}^2 \, {\cal R}_{27}^2 \, , \\[2.5mm]
  M_2 &=& \rho^4\, \mu_1 \, \mu_2 \, \mu_3^2 \, \mu_5^2 \, \mu_7  \,
     {\cal D}_{47}^2  \, {\cal D}_{57}^2 \, {\cal D}_{67}^2
  \, {\cal R}_{12}^2 \, {\cal R}_{13}^2 \, {\cal R}_{14}^2 \, {\cal R}_{16}^2 \, , \\[2.5mm]
  M_3 &=& \rho^4\, \mu_1^3 \, \mu_2^2 \, \mu_3^3 \, \mu_4 \, \mu_6  \,
    {\cal D}_{47}^2  \, {\cal D}_{57}^2 \, {\cal D}_{67}^2
  \, {\cal R}_{15}^4 \, {\cal R}_{17}^2 \, .
\end{eqnarray}
Here the parameters, $a_i\ (1,\cdots,7)$ are ordered as $a_1<a_2<a_3<a_4<a_5<a_6<a_7
$.
Let us introduce a scale $L=\sqrt{a_7-a_1}$ and the dimensionless parameters defined by 
\begin{eqnarray}
\kappa_i=\frac{a_{i+1}-a_1}{L^2},
\end{eqnarray}
and then these turn out to satisfy the inequalities
\begin{eqnarray}
0 < \kappa_1< \kappa_2< \kappa_3< \kappa_4< \kappa_5< 1.
\end{eqnarray}

\noindent
(i){\sl  Regularity}: From regularity, to get rid of singularities which appear at $z=a_1$ and $z=a_7$ on the rod $\rho=0$, the parameters should be set to be
\begin{eqnarray}
c_1  =
  -\frac{L\sqrt{\;2 \; \kappa_1 \; \kappa_2 \; \kappa_3 \; \kappa_5}}
  {\kappa_4},\ 
  b_2=  -L\;(1-\kappa_2)\sqrt{  \frac{2\;(1-\kappa_1)}
     {(1-\kappa_3) \;(1- \kappa_4) \; (1-\kappa_5)}} \;.
\end{eqnarray}
After imposing this condition on the parameters, the rod structure of the bicycling black rings can be summarized as follows:
\begin{itemize}
\item The semi-infinite spacelike rod $(-\infty,\kappa_1]$ with the rod vector $v=(0,1,0)$. To avoid conical singularities on $\rho=0,\ \bar z\in(-\infty,\kappa_1] $, i.e., to assure asymptotic flatness,  the angular coordinate $\phi$ should the period $\triangle \phi=2\pi$.
\item The finite timelike rod $[\kappa_1, \kappa_2]$ with the rod vector $v=(1,\Omega_\phi^{(1)},\Omega_\psi^{(1)})$ corresponds to a black ring whose
  horizon has the $S^1$ parameterized by the angular coordinate $\psi$ and the $S^2$ by the coordinates $(z,\phi)$, where $\Omega_\phi^{(1)}$ and $\Omega_\psi^{(1)}$ denote the angular velocities along the $\phi$-direction and the $\psi$-direction of the  rotating black ring, respectively, which are written as
\begin{eqnarray}
\Omega^{(1)}_\phi  = \frac{1}{L\, (1-\kappa_2)}
  \sqrt{\frac{(1-\kappa_3)(1-\kappa_4)(1-\kappa_5)}{2\, (1-\kappa_1)}} \, ,
\quad
  \Omega^{(1)}_\psi  = \frac{\kappa_4}{L}
  \sqrt{\frac{\kappa_1}{2\, \kappa_2\,\kappa_3\,\kappa_5}}.\nonumber
\end{eqnarray}
  
\item The finite spacelike rod $[\kappa_2,\kappa_3]$ with the rod vector $(0,1,0)$.

\item The finite spacelike rod $[\kappa_3,\kappa_4]$ with the rod vector $(0,0,1)$.

\item The finite timelike rod $[\kappa_4, \kappa_5]$ with the rod vector $v=(1,\Omega_\phi^{(1)},\Omega_\psi^{(1)})$ corresponds to a black ring whose
  horizon has the $S^1$ parameterized by the angular coordinate $\phi$ and the $S^2$ by the coordinates $(z,\psi)$, where $\Omega_\phi^{(2)}$ and $\Omega_\psi^{(2)}$ denote the angular velocities along the $\phi$-direction and the $\psi$-direction of the  rotating black ring, respectively. They are written as
  \begin{eqnarray}
&&\Omega^{(2)}_\phi = \frac{1-\kappa_2}{L}
  \sqrt{\frac{(1-\kappa_5)}{2\, (1-\kappa_1)(1-\kappa_3)(1-\kappa_4)}} \, ,
\quad  \Omega^{(2)}_\psi  =  \frac{1}{L\, \kappa_4}
  \sqrt{\frac{\kappa_1 \,\kappa_2 \, \kappa_3}{2\,  \kappa_5}}.\nonumber
\end{eqnarray}

\item The semi-infinite spacelike rod $[\kappa_5,\infty)$ with the rod vector $v=(0,0,1)$. To avoid conical singularities on $\rho=0,\ \bar z\in [\kappa_5,\infty)$, i.e., to assure asymptotic flatness,  the angular coordinate $\psi$ should the period $\triangle \psi=2\pi$.

\end{itemize}

\noindent
(ii) {\sl  Balance condition}: The absence of conical singularities on the rods $\rho=0, \ \bar z \in [\kappa_2,\kappa_3]$ and $\rho=0,\ \bar z\in[\kappa_3,\kappa_4]$, which correspond to the inner $\phi$-axis and $\psi$-axis between two orthogonal black rings,  requires the following conditions, respectively
\begin{eqnarray}
  1 ~=~ \frac{\Delta\phi}{2\pi} &=& \frac{\sqrt{\kappa_3\kappa_5(1-\kappa_1)
    (\kappa_3-\kappa_2)(\kappa_4-\kappa_1)(\kappa_4-\kappa_2)
    (\kappa_5-\kappa_2)}}{\kappa_4(1-\kappa_2)(\kappa_3-\kappa_1)
    (\kappa_5-\kappa_1)}\, ,  \cr
 1~=~\frac{\Delta\psi}{2\pi} &=&\frac{\sqrt{\kappa_5(1-\kappa_1)(1-\kappa_3)
    (\kappa_4-\kappa_1)(\kappa_4-\kappa_3)(\kappa_4-\kappa_2)
    (\kappa_5-\kappa_2)}}{\kappa_4(1-\kappa_2)(\kappa_5-\kappa_1)
    (\kappa_5-\kappa_3)}\, .
\end{eqnarray}
These are interpreted as the conditions that the two black rings should keep balance and prevent each other from collapsing by gravitational attraction. In conclusion, to summarize, from the requirement for the absence of singularities and conical singularities,  the total number of independent parameters for the balanced bicycling black ring amounts to be four, which means degrees of freedom of each mass, each angular momentum along the $S^1$ direction (one has the Komar angular momentum in the $\psi$-direction and the other has that in the $\phi$-direction) for two black rings.

\medskip
Finally, we briefly summarize the physical properties of the balanced bicycling black ring solution, following Ref.~\citen{Elvang-R}.
\begin{itemize}
\item {\sl Continuous non-uniqueness}: The balanced bicycling black rings exhibits $1$-fold non-uniqueness.
\item {\sl Frame-dragging}: As the black saturn does, two black objects interact with each other gravitationally. The spin along the $S^1$ direction of one black ring make the $S^2$ of the other black ring rotate, so that unlike the black saturn, each black ring has two non-vanishing angular velocities by the effect of frame-dragging, while each black ring has vanishing angular momentum for the $S^2$-rotation. 
\end{itemize}

\subsubsection{Other multiple black holes}
In four-dimensional Einstein theory, the multi-Kerr black hole solution is unlikely to exist.
In five dimensions, this is not trivial.
In analogy with such multi-black holes,  
Tan and Teo~\cite{TanTeo} first considered the configuration of static, asymptotically flat multi-black holes whose topology is $S^3$ as a solution of the five-dimensional Einstein equation.
They constructed the solution within a class of generalized Weyl solutions but such a  solution yields conical singularities between each black hole. There cannot exist such regular multi-black hole solutions within a class of asymptotically flat, static, vacuum solution due to the uniqueness theorem for static black holes~\cite{}. However, in five dimensional rotational case, this is not trivial since the spin-spin interaction between black holes may be so strong that it can balance gravitational attraction and prevent each black holes from collapsing one another. Recently, the work along this line was generalized to the rotational case~\cite{multi-MP}, namely, multi-Myers-Perry black hole solution with a single angular momentum was considered to construct. However, as shown in Ref.~\citen{multi-MP}, even for the rotational solution, the existence of conical singularities seems to be unable to be avoided.

\section{Kaluza-Klein black holes}\label{sec:KK}

A non-trivial class of exact solutions of Kaluza-Klein black holes is given by
{\sl Squashed} Kaluza-Klein black holes, where the squashing technique was applied to
five-dimensional black holes. 
The squashing transformation, which we will focus on later, has been then recognized as a type of solution 
generating technique. 
The idea is that for, e.g., the simplest  
static vacuum case, one first views the $S^3$ section 
(or horizon manifold) of a five-dimensional Schwarzschild-type black hole 
spacetime as a fiber bundle of $S^1$ over $S^2$, and then considers 
a deformation that changes the ratio of the radius of the fiber $S^1$ 
and base $S^2$, so that the resultant spacetime looks, 
at large distances, like a twisted $S^1$ over a four-dimensional asymptotically
flat spacetime, hence a Kaluza-Klein spacetime, while it looks like 
a five-dimensional black hole near the event horizon.
A recent key work along this line is that of Ishihara and Matsuno~\cite{IM}, 
who have found static charged Kaluza-Klein black hole solutions in the 
five-dimensional Einstein-Maxwell theory by using, for the first time, 
the squashing technique. 
Namely, they view the base manifold $S^3$ of a five-dimensional 
Reissner-Nordst\"om black hole as a fiber bundle over $S^2$ with 
fiber $S^1$ and then consider a deformation that changes the ratio of 
the radius of $S^2$ to that of $S^1$ fiber. It was shown by Wang~\cite{Wang} 
that the five-dimensional Kaluza-Klein black hole of Dobiasch and Maison can 
be reproduced by squashing a five-dimensional Myers-Perry black hole 
with two equal angular momenta. 
Subsequently the Ishihara-Matsuno solution was 
generalized to many different cases. For example, a static multi Kaluza-Klein 
black hole solution~\cite{IKMT} was constructed immediately from 
the Ishihara-Matsuno solution.  
A number of further generalizations of squashed Kaluza-Klein black
holes has been made lately~\cite{NIMT,TIMN,MINT,TI,T} in the five-dimensional supergravity theories. 
See, for example, refs.~\citen{Gaiotto,Elvang3,Gauntlett} for the earlier works in supergravities.

\medskip
As far as the present authors know, the squashing transformation 
has so far been applied only to cohomogeneity-one black hole solutions 
such as five-dimensional Myers-Perry black hole solutions with two equal 
angular momenta and the Cveti{\v c}-Youm's charged black hole solutions  
with two equal angular momenta.  It would be interesting to consider a generalization of the squashing technique which is applicable to cohomogeneity-two spacetimes 
such as black ring~\cite{Emparan:2001wn,Pom}, black holes with two 
rotations~\cite{Myers:1986un}, black lens~\cite{LMP,LMP2}, or even wider class 
of spacetimes.
 See Refs.~\cite{Rasheed,TYM,GS} for a cohomogeneity-two class of Kaluza-Klein 
solutions, which was found by utilizing another solution-generation-technique, e.g., a transformation by a non-linear sigma model.

\subsection{Kaluza-Klein black hole solutions in Einstein theories}\label{sec:rasheed}

\subsubsection{Kaluza-Klein black hole solutions in five dimensions}

Let us start from the action of the five-dimensional Einstein theory, which is given by
\begin{eqnarray}
S=\frac{1}{16\pi G}\int d^5x \sqrt{-g}R, \label{eq:action}
\end{eqnarray}
where $G$ is the five-dimensional Newton constant.
The five-dimensional metric can be written
\begin{eqnarray}
ds^2=e^{\frac{4\sigma}{\sqrt{3}}}(dx^5+2A_\mu dx^\mu)^2+e^{-\frac{2\sigma}{\sqrt{3}}}g_{\mu\nu}dx^\mu dx^\nu,
\end{eqnarray}
where $g_{\mu\nu}$ ($\mu,\nu=0,\cdots, 3$) is a four-dimensional metric.
The fifth coordinate $x^5$ is assumed to have with period $2\pi L$, and furthermore, the size of the fifth dimension, $L$, is assumed to be sufficiently small.  The low energy theory can be reduced by considering that the vector $\partial_{x^5}$ is a Killing vector of the five-dimensional spacetime so that all metric components, $g_{\mu\nu}$, $A_\mu$ and $\sigma$, do not depend on the coordinate $x^5$. The action (\ref{eq:action}), when dimensionally reduced to four dimensions,  becomes
\begin{eqnarray}
S=\int dx^4\sqrt{-g}\left[R-2(\partial\sigma)^2+e^{2\sqrt{3}\sigma}F^2\right],\label{eq:4-action}
\end{eqnarray}
where $F_{\mu\nu}=\partial_\mu A_\nu-\partial_\nu A_\mu$. Thus, the effective low energy theory turns out to be the four-dimensional gravity plus a Maxwell field and a scalar field.

\medskip
Let us consider the simplest example such that the fifth dimension and the four-dimensional spacetime is a trivial bundle:
\begin{eqnarray}
ds^2=(dx^5)^2+g_{\mu\nu}dx^\mu dx^\nu,
\end{eqnarray}
where the four-dimensional metric is a vacuum solution of the four-dimensional Einstein equation.
We evidently see that this five-dimensional metric satisfies the five-dimensional Einstein equation. If the four-dimensional spacetime describes a black hole, the five-dimensional solution is called a black string solution, and if they are a twisted non-trivial bundle at infinity, we here call it black hole or black ring, according as the spatial topology of the cross section of the horizon is $S^3$ or $S^1\times S^2$. Now let us consider the Lorentz boost in the direction of the fifth dimension, 
\begin{eqnarray}
t \to \cosh\alpha \ t+\sinh \alpha\  x^5,\quad x^5 \to \sinh\alpha\ t+\cosh\alpha \ x^5. 
\end{eqnarray}
Then under the transformation, the five-dimensional metric is also a solution of the five-dimensional Einstein equation since this is simply a coordinate transformation. The four-dimensional metric is, however, no longer a solution of the four-dimensional Einstein equation but a non-trivial solution of the equations derived from the four-dimensional action~(\ref{eq:4-action}). Therefore, from the point of view of four dimensions, this transformation yields a non-trivial electric charge, which is called a {\it  Kaluza-Klein (K-K) electric charge}. The spatial twist of the four-dimensional metric and the fifth dimension can yield a magnetic charge, which we call {\it Kaluza-Klein magnetic monopole}.

\medskip
In the five-dimensional Einstein theory, the black hole solution which has a K-K electric charge but no K-K magnetic charge, i.e., the boosted Schwarzschild string solution, was studied by Chodos and Detweiler~\cite{Chodos-Detweiler}, and then immediately this solution was generalized to the rotational case by Frolov, Zel'nikov and Bleyler~\cite{FZB}.
 %and Belinski-Ruffini~\cite{BR}.
 As mentioned previously, the first non-trivial solution, by which we means that the Kerr black hole and a Kaluza-Klein circle is twisted, was given by Dobiasch and Maison~\cite{DM}, who used a transformation by a non-linear sigma model (Recently, this Dobiasch-Maison solution was re-derived by squashing   from the five-dimensional Myer-Perry with equal angular momenta~\cite{Wang}.).
The Dobiasch-Maison solution was studied in detail by Gibbons-Wilshire~\cite{GW}, Pollard~\cite{Pollard} and Gibbons and Maeda~\cite{Gibbons-Maeda}. 

\medskip

Furthermore, the most general rotating Kaluza-Klein black hole solution in the five-dimensional Einstein gravity which has both a K-K electric charge and a K-K magnetic charge (i.e., a rotating dyonic black hole in the sense of four dimensions) was constructed by Rasheed~\cite{Rasheed}, who derived the black hole solution by applying the $SL(3,R)$ transformation to the Kerr string with the mass parameter $M_k$ and rotation parameter $a$. As shown by Maison~\cite{Maison}, the five-dimensional Einstein equation can be derived from the action describing the three-dimensional gravity-coupled
sigma model with five scalar fields, which is invariant under a global $SL(3,R)$ transformation. 
In order to assure the asymptotic flatness in the direction of four dimensions, the $SL(3,R)$ transformation must be restricted to the special $SO(1,2)$ transformation labeled by two boost parameters $(\alpha,\beta)$. 
Thus, the generated new Kaluza-Klein black hole solution is specified by the four parameters, $(M_k,a,\alpha,\beta)$.
The metric is given by
\begin{eqnarray}
ds^2&=&\frac{B}{A}(dx^5+2A_\mu dx^\mu)^2-\frac{f^2}{B}(dt+\omega^0{}_\phi d\phi)^2\nonumber\\
    & &+\frac{A}{\Delta}dr^2+Ad\theta^2+\frac{A\Delta}{f^2}\sin^2\theta d\phi^2,\label{eq:Rasheed}
\end{eqnarray}
where
\begin{eqnarray}
&&A=\left(r-\frac{\Sigma}{\sqrt{3}}\right)^2-\frac{2P^2\Sigma}{\Sigma-\sqrt{3}M}+a^2\cos^2\theta+\frac{2JPQ\cos\theta}{(M+\Sigma/\sqrt{3})^2-Q^2},\\
&&B=\left(r+\frac{\Sigma}{\sqrt{3}}\right)^2-\frac{2Q^2\Sigma}{\Sigma+\sqrt{3}M}+a^2\cos^2\theta-\frac{2JPQ\cos\theta}{(M-\Sigma/\sqrt{3})^2-P^2},\\
&&C=2Q(r-\Sigma/\sqrt{3})-\frac{2PJ\cos\theta(M+\Sigma/\sqrt{3})}{(M-\Sigma/\sqrt{3})^2-P^2},\\
&&\omega^0{}_\phi=\frac{2J\sin^2\theta}{f^2}\left[r-M+\frac{(M^2+\Sigma^2-P^2-Q^2)(M+\Sigma/\sqrt{3})}{(M+\Sigma/\sqrt{3})^2-Q^2}\right],\\
&&\omega^5{}_\phi=\frac{2P\Delta \cos\theta}{f^2}-\frac{2QJ\sin^2\theta[r(M-\Sigma/\sqrt{3})+M\Sigma/\sqrt{3}+\Sigma^2-P^2-Q^2]}{f^2[(M+\Sigma/\sqrt{3})^2-Q^2]},\\
&&\Delta=r^2-2Mr+P^2+Q^2-\Sigma^2+a^2,\\
&&f^2=r^2-2Mr +P^2+Q^2-\Sigma^2+a^2\cos^2\theta,\\
&&2A_\mu dx^\mu=\frac{C}{B}dt+\left(\omega^5{}_\phi+\frac{C}{B}\omega^0{}_\phi\right)d\phi,
\end{eqnarray}
where $A_\mu$ describes the electromagnetic vector potential derived by dimensional reduction to four dimension. Here the constants, $(M,P,Q,J,\Sigma)$, mean the mass, Kaluza-Klein magnetic charge (so-called NUT charge), Kaluza-Klein electric charge, angular momentum along four dimension and dilaton charge, respectively, which are parameterized by the two boost parameters $(\alpha,\beta)$
\begin{eqnarray}
&&M=\frac{(1+\cosh^2\alpha\cosh^2\beta)\cosh\alpha}{2\sqrt{1+\sinh^2\alpha\cosh^2\beta}}M_k,\\
&&\Sigma=\frac{\sqrt{3}\cosh\alpha(1-\cosh^2\beta+\sinh^2\alpha\cosh^2\beta)}{2\sqrt{1+\sinh^2\alpha\cosh^2\beta}}M_k,\\
&&Q=\sinh\alpha\sqrt{1+\sinh^2\alpha\cosh^2\beta}\ M_k,\\
&&P=\frac{\sinh\beta \cosh\beta}{\sqrt{1+\sinh^2\alpha\cosh^2\beta}}M_k,\\
&&J= \cosh\beta \sqrt{1+\sinh^2\alpha\cosh^2\beta\ }aM_k.
\end{eqnarray}
Note that all the above parameters are not independent since they are related through the equation
\begin{eqnarray}
\frac{Q^2}{\Sigma+\sqrt{3}M}+\frac{P^2}{\Sigma-\sqrt{3}M}=\frac{2\Sigma}{3},
\end{eqnarray}
and the constant $M_k$ is written in terms of these parameters
\begin{eqnarray}
M_k^2=M^2+\Sigma^2-P^2-Q^2.
\end{eqnarray}
The constant $J$ is also related to $a$ by
\begin{eqnarray}
J^2=a^2\frac{[(M+\Sigma/\sqrt{3})^2-Q^2][(M-\Sigma/\sqrt{3})^2-P^2]}{M^2+\Sigma^2-P^2-Q^2}.
\end{eqnarray}
The horizons exist at the values of $r$ which satisfies the quadratic equation $\Delta=0$ when the parameters satisfy the inequality
\begin{eqnarray}
M^2\ge P^2+Q^2+a^2-\Sigma^2.
\end{eqnarray}
From a four-dimensional point of view, the Rasheed solution describes the most general rotating dyonic black hole (in the five-dimensional Einstein theory), and can be specified by the four parameters $(M,J,Q,P)$ instead of the parameters $(M_k,a,\alpha,\beta)$ and includes the earlier known five-dimensional Kaluza-Klein black hole solutions. From Table~\ref{5dKKBH} the reader can see the relation between the five-dimensional Kaluza-Klein black hole solutions.

\begin{table}
\begin{center}
\begin{tabular}{l|c|c|c|c} \hline
 & $M$ & $J$ & $Q$ & $P$ \\ \hline
 Chodos-Detweiler (1982) & yes & no & yes & no \\ %\hline
 Frolov-Zel'nikov-Bleyer (1987) & yes & yes & yes & no \\ %\hline
 Dobiasch-Maison (1982) & yes & no & yes & yes \\ %\hline
 Rasheed (1995) & yes & yes & yes & yes \\ \hline
\end{tabular}
\caption[Classification of Kaluza-Klein black holes in the five-dimensional Einstein theory]{Classification of Kaluza-Klein black holes in the five-dimensional Einstein theory}
\label{5dKKBH}
\end{center}
\end{table}

\medskip
\subsubsection{Kaluza-Klein black holes in higher dimensions}
The above type of Kaluza-Klein solutions were further generalized to  higher dimensions ($D\ge 6$) with $U(1)^n\ (n\ge 2) $ internal isometries ~\cite{CY-KK,CY-KK2}, 
and the general black holes are parameterized by its mass, $n$ Kaluza-Klein electric charges and $n$ Kaluza-Klein magnetic monopole charges~\cite{CY-KK2}.

\subsection{Charged Kaluza-Klein black hole solutions in supergravity}\label{sec:IM}
Kaluza-Klein black hole solutions which asymptote to the non-trivial bundle 
of the four-dimensional Minkowski space-time and $S^1$ were generalized in the context of supersymmetric solutions.
The first supersymmetric Kaluza-Klein black hole solution of this type was constructed 
as a black hole in Taub-NUT space~\cite{Gauntlett,Gaiotto}.
A similar type of supersymmetric black hole  was obtained in Ref.~\citen{Elvang3} by taking the limit of a supersymmetric black ring in Taub-NUT space to a black hole but 
as discussed in Ref.~\citen{Elvang3}, these solutions differ from each other only in the existence of a magnetic charge from the four-dimensional point of view. 
The further generalizations to another supergravity, or black strings, or black rings in Taub-NUT space were also considered by many authors~\cite{BKW,Bena,Bena2,Bena3,BGRW,Elvang3,EEMR2,FGPS,CEFGS,BDGRW,GRS,CBJV,T,Gibbons-Perry,CY-KK3,CY-KK4,CY-KK5,Nelson}.

\medskip
For non-extremal cases, a similar type of Kaluza-Klein black holes was considered 
by Ishihara-Matsuno~\cite{IM}, who have found static charged Kaluza-Klein black hole solutions in the 
five-dimensional Einstein-Maxwell theory by using the squashing technique for the five-dimensional Reissner-Nordstr\"om solution. 
Subsequently the Ishihara-Matsuno solution was generalized to many different cases in the five-dimensional supergravity theories. 
In Ref. \citen{NIMT}, the squashing transformation was applied 
to the five-dimensional Cveti{\v c}-Youm charged rotating black hole 
solution~\cite{CY96} with equal charges and then as a result, an non-extremal charged rotating Kaluza-Klein black hole solution with the supersymmetric limit~\cite{Gauntlett,Gaiotto} was obtained.  
Furthermore, the application of the squashing transformation to non-asymptotically 
flat Kerr-G\"odel black hole solutions~\cite{Gimon-Hashimoto,Herdeiro,Wu} was first 
considered in Refs. \citen{TIMN, MINT, TI}. Remarkably, although the G\"odel black hole solution as a seed has closed timelike curves in the far region from the black hole, the black hole solution obtained in the way described above have no closed timelike curve in the domain of outer communication. Immediately, this type of solutions was generalized to multi-black hole solutions~\cite{IKMT,MINT}. 
These solutions \cite{NIMT,TIMN,TI} correspond to 
a generalization of the Ishihara-Matsuno solution to the rotating 
black holes in the bosonic sector of the five-dimensional minimal supergravity, i.e., in Einstein-Maxwell-Chern-Simons theory with a certain coupling. See Table~\ref{squashing} about the relation between the non-compactified solutions and the corresponding squashed (compactified) solutions in the five-dimensional Einstein gravity, or supergravity.  

\medskip
As mentioned previously, this squashing transformation deforms a cohomogeneity-one class of solutions 
such as five-dimensional rotating black holes with equal angular momenta into a cohomogeneity-one class of Kaluza-Klein type solutions.
Note that rotating black hole solutions with unequal rotations~\cite{Myers:1986un}, black rings~\cite{Emparan:2001wn,Pom}, or black lens~\cite{LMP,LMP2} belong to a cohomogeneity-two class of solutions. Therefore, cohomogeneity-two Kaluza-Klein black hole solutions such as the Rasheed solution~\cite{Rasheed} cannot be obtained by using this method. 
Recently, this Rasheed solution was generalized to charged cases in the five-dimensional minimal supergravity~\cite{TYM,GS} by utilizing the sigma-model technique for minimal five-dimensional supergravity~\cite{Galtsov,Galtsov2}.

\begin{table}
\begin{center}
\begin{tabular}{l|l}\hline
Non-compactified solutions %in $D=5$ Eisntein gravity/supergravity  
& Corresponding squashed solutions 
\\ \hline
Minkowski & Gross-Perry-Sorkin monopole (1983) \\ %\hline
Myers-Perry solution with equal angular momenta & Dobiasch-Maison solution (1982)\\ %\hline
BMPV solution & Gaiotto-Strominger-Yin solution (2006)\\ %\hline
 Reissner-Nordstr\"om solution & Ishihara-Matsuno (2006) \\ %\hline
Cveti{\v c}-Youm solution with equal charges & Nakagawa-Ishihara-Matsuno-Tomizawa (2008) \\ %\hline
G\"odel universe & rotating GPS monopole (2009) \\ %\hline
Kerr-Newman-G\"odel solution & Tomizawa-Ishihara-Matsuno-Nakagawa (2009) \\ %\hline
Reissner-Nordstr\"om-G\"odel solution & Tomizawa-Ishibashi (2008) \\ %\hline
Cveti{\v c}-Youm solution & Tomizawa (2010) \\ \hline
\end{tabular}
\caption[smallcaption]{The relation between a cohomogeneity-one class of non-compactified solutions and squashed solutions: The five-dimensional solutions of Einstein theory, 
or supergravity in the right hand side can be obtained by squashing transformation from the corresponding solutions of the same theory in the left hand side.}
\label{squashing}
\end{center}
\end{table}

\medskip 
Let us consider Kaluza-Klein black hole solutions in the bosonic sector of the five-dimensional minimal supergravity, 
i.e., the five-dimensional Einstein-Maxwell-Chern-Simons theory with a certain coupling, whose action is given by  
\begin{eqnarray}
 S = \frac{1}{16 \pi G_5} 
    \int d^5 x \left[ 
                    \sqrt{-g} \left( R - F_{MN } F^{MN } \right) 
                    + \frac{2}{3 \sqrt 3} 
                      \epsilon ^{MN KLP } 
                      A_M F_{N K } F_{LP } 
               \right] \,, 
\label{action}
\end{eqnarray}
where $R$ is the five dimensional scalar curvature, 
$F = dA$ is the field strength of the gauge potential one-form $A$, and $G_5$ is 
the five-dimensional Newton constant. Varying the action (\ref{action}), 
we derive the Einstein equation 
\begin{eqnarray}
 R_{MN } -\frac{1}{2} R g_{MN } 
 = 2 \left( F_{M L } F_N^{ ~L } 
  - \frac{1}{4} g_{MN } F_{KL } F^{KL } \right) \,, 
 \label{Eineq}
\end{eqnarray}
and the Maxwell-Chern-Simons equation 
\begin{eqnarray}
 F^{MN}_{~~~; N} + \frac{1}{2 \sqrt 3} \left( \sqrt{-g} \right)^{-1} 
   \epsilon ^{MNKLP } F_{NK } F_{LP } 
   = 0 \,. 
 \label{Maxeq}  
\end{eqnarray}

\subsubsection{Static electrically charged solution}
The static electrically charged Kaluza-Klein black hole solution to the equations~(\ref{Eineq}) and (\ref{Maxeq}) was obtained by the squashing from the five-dimensional asymptotically flat Reissner-Nordstr\"om solution~\cite{IM}.
The metric is written
\begin{eqnarray}
ds^2=-fdt^2+\frac{k^2}{f}dr^2+\frac{r^2}{4}\left[k\left\{(\sigma^1)^2+(\sigma^2)^2\right\}+(\sigma^3)^2\right],\label{eq:IM-metric}
\end{eqnarray}
where $f$ is a function of $r$ defined by
\begin{eqnarray}
f(r)=\frac{(r^2-r_+^2)(r^2-r_-^2)}{r^4}.
\end{eqnarray}
$k(r)$---called the {\sl squashing function}---is given by 
\begin{eqnarray}
\quad k(r)=\frac{(r_\infty^2-r_+^2)(r_\infty^2-r_-^2)}{(r_\infty^2-r^2)^2},
\end{eqnarray}
and the gauge potential is
\begin{eqnarray}
A=\pm\frac{\sqrt{3}}{2}\frac{r_+r_-}{r^2}dt.\label{eq:IM-gauge}
\end{eqnarray}
Here, $r_\pm$
and $r_\infty$ are constants, and  $\sigma^i (i = 1, 2, 3)$ are $SU(2)$ invariant 1-forms satisfying the relation
\begin{eqnarray}
d\sigma^i=\frac{1}{2}C_{jk}^i\sigma^j\wedge\sigma^k
\end{eqnarray}
with $C^1_{23}=C^2_{31}=C^3_{12}=1$, $C^i_{jk}=0$ in all other cases. In terms of a coordinate basis, they are explicitly written as
\begin{eqnarray}
&&\sigma^1=\cos\psi d\theta+\sin\psi\sin\theta d\phi,\\
&&\sigma^2=-\sin\psi d\theta+\cos\psi\sin\theta d\phi,\\
&&\sigma^3=d\psi+\cos\theta d\phi
\end{eqnarray}
with the coordinates $(\theta,\phi, \psi)$ having the ranges,
 $0\le \theta<\pi$, $0\le \phi<2\pi$, $0\le \psi<4\pi$. The metric has horizons at $r = r_+$ (the outer horizon) and at $r = r_-$ (the inner horizon). 
Because the metric is apparently singular at $r = r_\infty$, the radial coordinate $r$ should be assumed to move the range $0<r<r_\infty$.
Note here that 
\begin{eqnarray}
&&\frac{1}{4}\left[(\sigma^1)^2+(\sigma^2)^2+(\sigma^3)^2\right]=d\Omega_{S^3}^2,\\
&&(\sigma^1)^2+(\sigma^2)^2=d\Omega_{S^2}^2
\end{eqnarray}
hold.
 It is immediate to see that the metric of the five-dimensional Reissner-Nordstr\"om black hole solution is transformed as 
$dr\to k(r)dr$, $\sigma_1\to \sqrt{k(r)}\sigma_1$ and 
$\sigma_2 \to \sqrt{k(r)}\sigma_2$. 
 By this transformation, the metric of the unit round $S^3$ is 
deformed to the metric of a squashed $S^3$ for which the radius 
of $S^2$ is no longer equal to the radius of $S^1$. 
For this reason, this is called the squashing transformation. 
Taking the limit of $r_\infty\to \infty$ ($k(r)\to 1$) for the above solution (\ref{eq:IM-metric}) and (\ref{eq:IM-gauge}) recovers the five-dimensional Reissner-Nordstr\"om solution.
The point $r=r_\infty$ turns out to correspond to spatial infinity. 
In order to confirm this, now let us introduce the following new radial coordinate 
\begin{eqnarray}
\rho =\rho_0 \frac{ r^2}{r_\infty^2-r^2} \,, 
\label{def:rho:RotGPS}
\end{eqnarray}
so that $r_\infty\to \infty$ corresponds to $\rho \to \infty$. Here the constant $\rho_0$ is defined by
\begin{eqnarray}
\rho_0=\frac{f(r_\infty)r_\infty^2}{4}.
\end{eqnarray}
The new coordinate $\rho$ varies from 0 to ‡ when $r$ varies from $0$ to $r_\infty$. 
The metric (\ref{eq:IM-metric}) can be
rewritten in terms of $ \rho$ and $T =\sqrt{f(r_\infty)}\ t$ as
\begin{eqnarray}
ds^2=-{\cal V}dT^2+\frac{K^2}{{\cal V}}d\rho^2+R^2\{(\sigma^1)^2+(\sigma^2)^2\}+W^2(\sigma^3)^2,\label{eq:IM-metric2}
\end{eqnarray}
where
\begin{eqnarray}
{\cal V}=\frac{(\rho-\rho_+)(\rho-\rho_-)}{\rho^2},\quad K=1+\frac{\rho_0}{\rho},\quad R^2=\rho^2K^2,
\end{eqnarray}
\begin{eqnarray}
W^2=(\rho_0+\rho_+)(\rho_0+\rho_-)K^{-2}
\end{eqnarray}
with the constants $\rho_\pm$ defined by
\begin{eqnarray}
\rho_\pm=\rho_0\frac{r_\pm^2}{r^2_\infty-r^2_\pm}.
\end{eqnarray}
Near $\rho\to\infty$ ($r\to r_\infty$), the metric turns out to behave as
\begin{eqnarray}
ds^2\simeq -dT^2+d\rho^2+\rho^2(d\theta^2+\sin^2\theta d\phi^2)+\frac{r_\infty^2}{4}(d\psi+\cos\theta d\phi)^2.
\end{eqnarray}
It is now clear that the metric is asymptotically locally flat 
and has the structure of a twisted $S^1$ bundle over 
the four-dimensional Minkowski space-time.

\subsubsection{Kaluza-Klein multi-black holes}
The above solution can be generalized to solution with multiple horizons.
For a single extreme black hole, the metric and the gauge potential one-form, which can be obtained by 
setting the parameters and the coordinate to be $M:=\rho_+=\rho_-$, $N:=\rho_0+\rho_+$ and $x^5:=N\psi$ in (\ref{eq:IM-metric2}), are written as
\begin{eqnarray}
 ds^2 &=& - H^{-2} dT^2 + H ds^2_{\text{TN}},
        \label{SUSYBH} \\
    A &=& \pm \frac{\sqrt{3}}{2} H^{-1} dT.
\label{gaugepotintial}
\end{eqnarray}
When the Taub-NUT space is described in the form 
\begin{eqnarray}
 &&ds^2_{\text{TN}} 
		= V^{-1} \left( d\rho^2 + \rho^2d\Omega_{S^2}^2 \right)  
		+V \left( dx^5+ N\cos\theta d\phi\right) ^2, \label{TN} \\
 &&V^{-1} ( \rho) = 1 + \frac{ N }{ \rho },                                          \label{HarmonicW}
\end{eqnarray}
where
\begin{eqnarray}
		0\leq x^5 \leq 2\pi L ,
\label{period}
\end{eqnarray}
the function $H$ is given by
\begin{eqnarray}
 H(\rho) &=&  1 + \frac{ M }{ \rho }.                                                     \label{HarmonicH}
\end{eqnarray} 
Here $L, M$ and $N$ are positive constants. 
Regularity of the spacetime requires that 
the nut charge, $N$, and the asymptotic radius of $S^1$ along 
$x^5$, $L$, are related by 
\begin{equation}
 N = \frac{ L }{ 2 } n,
\end{equation} 
where $n$ is a natural number. When we generalize this single black hole solution (\ref{SUSYBH}) to the multi-black holes~\cite{Gauntlett,IKMT}, 
it is natural to generalize the Taub-NUT space 
to the Gibbons-Hawking space~\cite{GH} which has multi-nut singularities. 
The metric form of the Gibbons-Hawking space is 
\begin{eqnarray}
& & d{s}^2_{\rm GH}
	=V^{-1} \left( d{\bm x}\cdot d{\bm x}\right)
		+V\left(dx^5 + {\bm\omega} \right)^2, 
\\
& &V^{-1} =1+\sum_i\frac{N_i}{|\bm{x}-\bm{x}_i|},
\label{042802}
\end{eqnarray}
where $\bm{x}_i=(x_i,y_i,z_i)$ denotes position of the $i$-th 
nut singularity with nut charge $N_i$ 
in the three-dimensional Euclid space, and $\bm\omega$ satisfies 
\begin{eqnarray}
{\bm \nabla}\times {\bm \omega} = {\bm\nabla} V^{-1}.
\label{0428025}
\end{eqnarray}
We can write down a solution ${\bm \omega}$ explicitly as
\begin{eqnarray}
{\bm \omega}
	=\sum_i N_i 
			\frac{(z-z_i)}{|\bm{x}-\bm{x}_i|}~
			\frac{(x-x_i)dy -(y-y_i)dx}{(x-x_i)^2+(y-y_i)^2}.
\end{eqnarray}

If we assume the metric form with the Gibbons-Hawking space instead 
of the Taub-NUT space in \eqref{SUSYBH}, 
the Einstein equation and the Maxwell equation reduce to 
\begin{equation}
	\triangle_{\rm GH} H=0, 
\label{042801}
\end{equation}
where $\triangle_{\rm GH}$ is the Laplacian of the Gibbons-Hawking space.
In general, it is difficult to solve this equation, 
but if one assume $\partial/\partial x^5$ to be a Killing vector, 
as it is for the Gibbons-Hawking space,  
then Eq. (\ref{042801}) reduces to the Laplace equation in the 
three-dimensional Euclid space, 
\begin{equation}
	\triangle_{\text E} H=0. \label{042804}
\end{equation}

We take a solution with point sources to Eq. \eqref{042804} 
as a generalization of Eq. \eqref{HarmonicH}, 
and we have the final form of the metric 
\begin{eqnarray}
&&
	ds^2=-H^{-2}dT^2 + H ds_{\rm GH}^2, 
\label{eq:ekk}
\\ 
&&H=1+\sum_i\frac{M_i}{|\bm{x}-\bm{x}_i|},
\label{eq:harmonic}
\end{eqnarray}
where  $M_i$  are constants.\footnote{
For the special case $H=1/V$ , the metric (\ref{eq:ekk}) reduces to
the four dimensional Majumdar-Papapetrou multi-black holes 
with twisted constant $S^1$. 
In this special case, $M_i$ are also quantized as well 
as nut charges.
}

\medskip
The induced metric on an intersection of the $i$-th black hole horizon 
with a static time-slice is
\begin{eqnarray}
ds^2_{{\rm Horizon}}
	=\frac{L M_i n_i}{2}\biggl[
	\biggl(\frac{d\psi}{n_i}+\cos\theta d\phi \biggr)^2
		+d\Omega_{S^2}^2\biggr],
\end{eqnarray}
where 
$0\leq \psi=2x^5/L \leq 4\pi$.   
In the case of $n_i=1$, it is apparent that 
the $i$-th black hole is a round $S^3$. In the case of $n_i\ge 2$, however, the topological structure 
becomes a lens space $L(n_i;1)= S^3/{\mathbb Z}_{n_i}$. 

\medskip
Let us see the asymptotic behavior of the Kaluza-Klein 
multi-black hole in the neighborhood of the spatial infinity 
$\rho\rightarrow\infty$. 
The functions $H$, $V^{-1}$ and 
${\bm \omega}$ behave as
\begin{eqnarray}
&&H(\rho,\theta)
	\simeq 1+\frac{\sum_i M_i}{\rho}+O\biggl(\frac{1}{\rho^2}\biggr),
\\
&&V(\rho,\theta)^{-1}
	\simeq 1+\frac{\sum_i N_i}{\rho}+O\biggl(\frac{1}{\rho^2}\biggr),
\\
&&{\bm\omega}(\rho,\theta)
	\simeq \biggl(\sum_i N_i\biggr)\cos\theta d\phi+O\biggl(\frac{1}{\rho}\biggr).
\end{eqnarray}
We can see that the spatial infinity possesses the structure of 
$S^1$ bundle over $S^2$ such that it is a 
lens space $L(\sum_i n_i;1)$. 
For an example, in the case of two Kaluza-Klein black holes 
which have the same topological structure of $S^3$, 
the asymptotic structure is  topologically  homeomorphic 
to the lens space $L(2;1)=S^3/{\mathbb Z}_2$.  
{}From this behavior of the metric near the spatial infinity, 
we can compute the Komar mass at spatial infinity of this multi-black hole 
system as 
\begin{eqnarray}
	M_{\text{Komar}}=\frac{3\pi}{2G}L\sum_i M_i.
\end{eqnarray}
Since the total electric charge is given by
\begin{eqnarray}
	Q_{\text{total}}=\sum_i Q_i = \frac{\sqrt{3}\pi}{G}L\sum_i M_i, 
\end{eqnarray}
then the total Komar mass and the total electric charge satisfy
\begin{eqnarray}
	M_{\text{Komar}}= \frac{\sqrt{3}}{2} |Q_{\text{total}}|.
\end{eqnarray}
Therefore, we find that an observer located in the neighborhood of 
the spatial infinity feels as if there were a single 
Kaluza-Klein black hole with the point source with  
the parameter $M=\sum_i M_i$ and the nut charge 
$N={\sum_i N_i}$.

\subsubsection{Charged rotating Kaluza-Klein black holes}
The electrically charged solution above was immediately generalized to the rotating case~\cite{Wang,NIMT}. 
Applying the squashing transformation to the five-dimensional Cveti{\v c}-Youm solution~\cite{CY96} with equal charges in the bosonic sector of the five-dimensional minimal supergravity yielded 
the electrically charged, rotating Kaluza-Klein black hole solution in the same theory~\cite{NIMT}.
The metric and the gauge potential of the solution obtained after the squashing transformation are given by
\begin{eqnarray}
ds^2=-\frac{w}{h}dt^2+\frac{k^2}{w}dr^2+\frac{r^2}{4}\biggl[k\{(\sigma^1)^2+(\sigma^2)^2\}+h\{fdt+(\sigma^3)^2\}^2\biggr],
\end{eqnarray}
and 
\begin{eqnarray}
A=\frac{\sqrt{3}q}{2r^2}\left(dt-\frac{a}{2}\sigma ^3\right),
\end{eqnarray}
respectively, where the metric functions $w, h,f$ and $k$ are defined as
\begin{eqnarray}
&&w(r)=\frac{(r^2+q)^2-2(m+q)(r^2-a^2)}{r^4},\\
&&h(r)=1-\frac{a^2q^2}{r^6}+\frac{2a^2(m+q)}{r^4},\\
&&f(r)=-\frac{2a}{r^2h(r)}\left(\frac{2m+q}{r^2}-\frac{q^2}{r^4}\right),\\
&&k(r)=\frac{(r_\infty^2+q)^2-2(m+q)(r_\infty^2-a^2)}{(r_\infty^2-r^2)^2}.
\end{eqnarray}
In what follows the four parameters $(m,q,a,r_\infty)$ that specify the above solutions are
assumed to satisfy the following inequalities,
\begin{eqnarray}
&&m>0,\label{eq:0in2}\\
&&q^2+2(m+q)a^2>0,\label{eq:0in3}\\
&&(r_\infty^2+q)^2-2(m+q)(r_\infty^2-a^2)>0,\label{eq:0in3-2}\\
&&(m+q)(m-q-2a^2)>0\label{eq:0in4},\\
&&m+q>0\label{eq:0in1}.
\end{eqnarray}
These inequalities are the necessary and sufficient conditions for the
spacetime to admit two Killing horizons and no CTC outside the horizon.

\medskip
In the static electrically charged solution~\cite{IM}, the $m=\pm q$ case  
corresponds to a supersymmetric solution, i.e., the static black hole 
in Taub-NUT space~\cite{Gaiotto,Gauntlett}. 
However, in the above rotating solution~\cite{NIMT}, 
the two cases, $m=-q$ and $m=q$, describe different solutions due to 
the existence of a Chern-Simons term. For either $m=q+2a^2$ 
or $m=-q$, the outer and inner horizons degenerate, but 
only the case of $m=-q$ corresponds to a supersymmetric Kaluza-Klein black 
hole solution~\cite{Gaiotto,Gauntlett}.

\subsubsection{Squashed G\"odel black holes}
The application of the squashing transformation to non-asymptotically 
flat Kerr-G\"odel black hole solution~\cite{Gimon-Hashimoto} which has closed timelike curves in the far region from the black hole, was also
considered in Ref. \citen{TIMN}, and as seen later, remarkably, the resulting squashed spacetime turns out to have no closed timelike curve.
Before considering such black hole solution in the G\"odel universe background, let us start from seeing the properties of the squashed G\"odel universe.

\medskip
{\sl Rotating GPS monopole}: 
We shall briefly review the results of \cite{TIMN} concerning the 
{\sl rotating Gross-Perry-Sorkin (GPS) monopole}, which is one of the simplest 
solutions in five-dimensional Einstein-Maxwell-Chern-Simons theories.  
We begin with the five-dimensional G\"odel universe since the rotating GPS 
monopole solutions are obtained by applying the squashing transformation 
to the five-dimensional supersymmetric G\"odel universe.  
We then discuss some basic properties of the 
rotating GPS monopole, in particular, its asymptotic structure 
and the existence of an ergoregion.

Then, as a solution, we have the following metric and the gauge potential 
one-form, respectively,   
\begin{eqnarray}
ds^2= -(dt+jr^2\sigma^3)^2 + dr^2 
      + \frac{r^2}{4}\left\{(\sigma^1)^2+(\sigma^2)^2+(\sigma^3)^2\right\} \,
\label{eq:Godel}
\end{eqnarray}
and 
\begin{eqnarray}
A=\frac{\sqrt{3}}{2}jr^2\sigma^3,\label{eq:Godel2} \,, 
\end{eqnarray}
with the coordinates $r$ having the range 
$0<r<\infty$.
The parameter, $j$, is called the {\sl G\"odel parameter}. 
The norm of the Killing vector $\partial_\psi$
becomes negative in the region of $r>1/(2|j|)$ with the signature remaining 
Lorentzian, and therefore this solution admits closed timelike curves (CTCs), 
as the four-dimensional G\"odel universe does. 
The metric, Eq.~(\ref{eq:Godel}), is completely homogeneous as a 
five-dimensional spacetime, just like the four-dimensional G\"odel universe 
is so in the four-dimensional sense. While the four-dimensional G\"odel 
universe is a spacetime filled with a pressureless perfect fluid balanced 
with a negative cosmological constant, 
the five-dimensional G\"odel universe given above, Eqs.~(\ref{eq:Godel}) 
and (\ref{eq:Godel2}), is filled with a a configuration of gauge field.

\medskip 
The rotating GPS monopole solution is given by the following metric and 
the gauge potential one-form:  
\begin{eqnarray}
ds^2&=& -(dt+jr^2\sigma^3)^2 + k^2dr^2 
        + \frac{r^2}{4}\left[k\{(\sigma^1)^2+(\sigma^2)^2\}+(\sigma^3)^2\right] \,, 
\label{eq:rotaingGPS}
\\
A &=& \frac{\sqrt{3}}{2}jr^2\sigma^3 \,, 
\label{eq:rotaingGPS2}
\end{eqnarray}
where the squashing function $k(r)$ is given by 
\begin{eqnarray}
k(r)=\frac{r_\infty^4}{(r_\infty^2-r^2)^2} \,.
\end{eqnarray}
It is immediate to see that the metric of the G\"odel universe, 
Eq.~(\ref{eq:Godel}), is transformed as 
$dr\to k(r)dr$, $\sigma_1\to \sqrt{k(r)}\sigma_1$ and 
$\sigma_2 \to \sqrt{k(r)}\sigma_2$.

\medskip 

Now let us introduce the following new radial coordinate 
\begin{eqnarray}
\rho = \frac{r_\infty r^2}{2(r_\infty^2-r^2)} \,, 
\label{def:rho:RotGPS}
\end{eqnarray}
so that $r\to r_\infty$ corresponds to $\rho \to \infty$. 
Then, the metric and the gauge potential one-form can be rewritten, 
respectively, as 
\begin{eqnarray}
 ds^2&=&-\left\{
               dt + 4j\rho_0^2
               \left( 1+\frac{\rho_0}{\rho}\right)^{-1}\sigma^3 
         \right\}^2 
\nonumber\\
  && + \left(1+\frac{\rho_0}{\rho}\right)
       \left[d\rho^2+\rho^2\{(\sigma^1)^2+(\sigma^2)^2\}\right] 
     +\rho_0^2\left(1+\frac{\rho_0}{\rho}\right)^{-1}(\sigma^3)^2 \,,
\label{eq:GPS}  
\end{eqnarray} 
\begin{eqnarray} 
  A=2\sqrt{3}\rho_0^2j\left(1+\frac{\rho_0}{\rho}\right)^{-1}\sigma^3 \,,\label{eq:GPSA}
\end{eqnarray}
where we have also introduced the constant $\rho_0=r_\infty/2$.

\medskip 
In this spacetime, depending on the choice of parameter range, 
we suffer from causality violation due to the existence of closed 
timelike curves (CTCs). 
(Recall that $\sigma^3$ in the first term of the above metric includes 
the periodic coordinate $\psi$.) 
To cure this, we hereafter restrict the range of the parameters 
$(j,\rho_0)$ as 
\begin{eqnarray}
j^2<\frac{1}{16\rho_0^2},\quad \rho_0>0 \,.
\label{rest:param:j-rho}
\end{eqnarray} 

\medskip 
When the G\"odel parameter is $j=0$, the metric, Eq.~(\ref{eq:GPS}), 
coincides with the static GPS monopole solution 
given originally in Ref.~\cite{GP}. In this case it is immediate to see that 
the point of $\rho=0$ is a fixed point of the Killing vector field, 
%$\psi^\mu =(\partial/\partial \psi)^\mu$, 
$\partial_\psi$, 
and the metric is analytic there, 
and thus the metric corresponds to a Kaluza-Klein monopole~\cite{GP}.  
A fixed point of some Killing field like this is often called a {\sl nut}. 
This is also the case even when $j \neq 0$.

\medskip 
Further, we introduce new coordinates defined by 
\begin{eqnarray}
\bar t=\frac{t}{C},\quad \bar\psi=\psi-\frac{D}{C}\ t \,,
\end{eqnarray}
where the constants $C$ and $D$ are 
\begin{eqnarray}
  C=\sqrt{1-16j^2\rho_0^2} \,,\quad 
  D=\frac{4j}{\sqrt{1-16j^2\rho_0^2}} \,,  
\end{eqnarray} 
which make sense under the condition, Eq.~(\ref{rest:param:j-rho}). 
For $\rho\to\infty$, the metric behaves as 
\begin{eqnarray}
ds^2 &\simeq& -d\bar t^2+d\rho^2+\rho^2\{(\sigma^1)^2+(\sigma^2)^2\}
              +\rho_0^2\left(1-16j^2\rho_0^2\right)(\sigma^3)^2 \,. 
\end{eqnarray} 
It is now clear that the metric is asymptotically locally flat 
and has the structure of a twisted $S^1$ bundle over 
the four-dimensional Minkowski space-time. 
It is also clear that the presence of non-vanishing parameter $j$
in Eq.~(\ref{eq:GPS}) means 
that 
the spacetime is 
{\sl rotating along the direction of the extra-dimension} 
specified by $\partial_\psi$.
We emphasize here again that CTCs which exist in the G\"odel universe, 
Eq.~(\ref{eq:Godel}), now cease to exist as a result of the squashing 
transformation and the choice of the parameter range, 
Eq.~(\ref{rest:param:j-rho}). 

\medskip 
Remarkably, this rotating GPS monopole solution possesses an ergoregion, 
despite the fact that there is no black hole event horizon in this spacetime. 
This can be seen as follows. 
The $\bar t\bar t$-component of the metric in the rest frame takes 
the following form near infinity,  
\begin{eqnarray}
 g_{\bar t\bar t} 
 = - \left[ 
          \left( 
                C+\frac{1-C^2}{C}\left(1+\frac{\rho_0}{\rho}\right)^{-1}
          \right)^2 
          -\frac{1-C^2}{C^2}\left(1+\frac{\rho_0}{\rho}\right)^{-1}
     \right] \,.
\end{eqnarray}
As shown in Fig.
%\ref{fig:GPS_ergo}, 
in the case of 
$\sqrt{3}/8<|j|\rho_0<1/4$, $g_{\bar t \bar t}$ becomes positive in 
the region of $\gamma_-<\rho<\gamma_+$, where 
\begin{eqnarray} 
\gamma_\pm:=\rho_0\frac{1-3C^2\pm(1-C^2)\sqrt{1-4C^2}}{2C^2} \,, 
\end{eqnarray}
and therefore there exists an ergoregion in that region, 
although there is no black hole horizon in the space-time. 
In a neighborhood of the nut, the ergoregion vanishes.

\if0
\begin{figure}[!h]
\begin{center}
\includegraphics[width=0.5\linewidth]{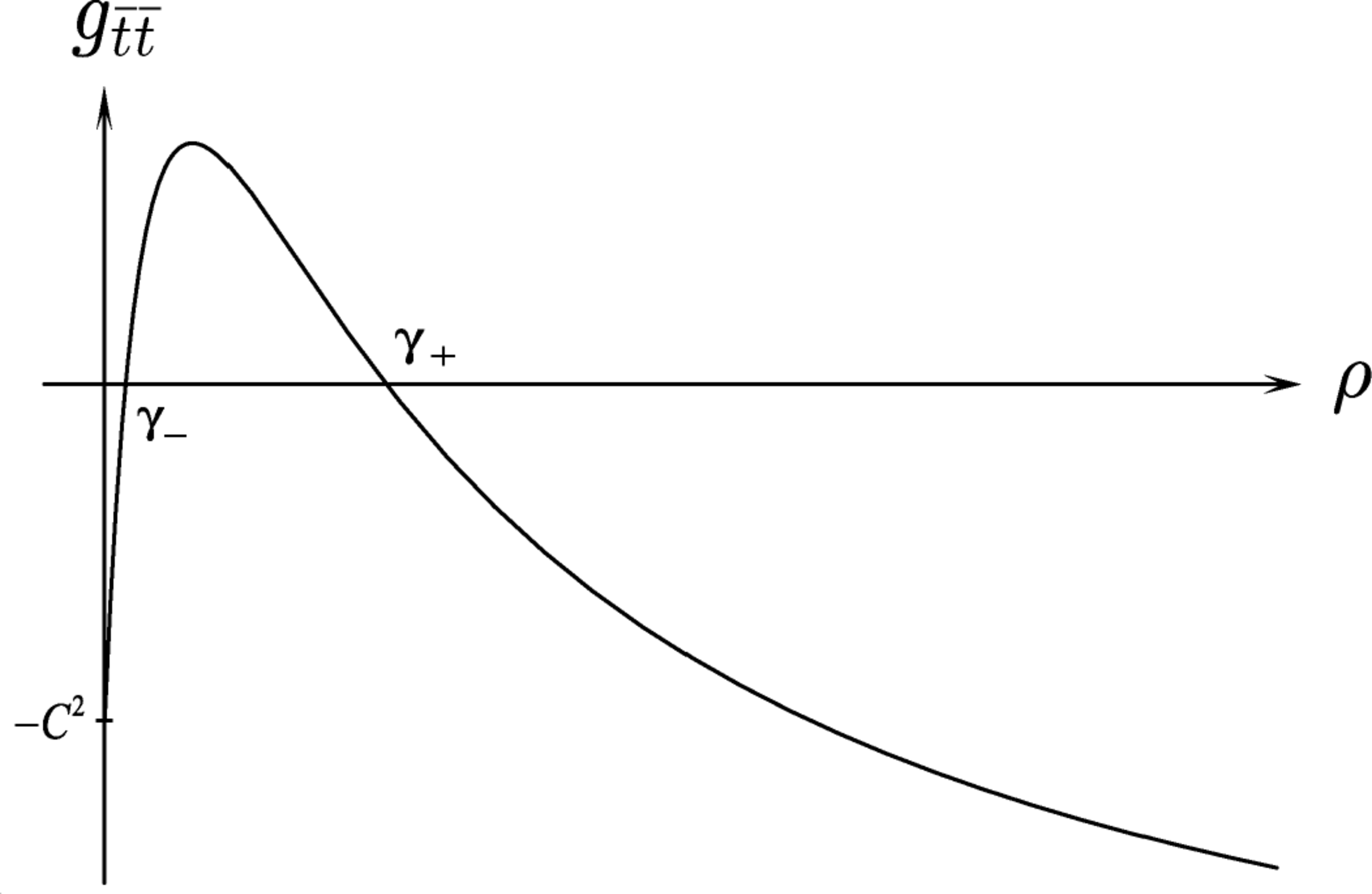}
\begin{minipage}{0.8\hsize}
\caption{The typical behavior of $g_{\bar t\bar t}$ in the case of 
$\sqrt{3}/8<|j|\rho_0<1/4$. There exists an ergoregion in the region 
such that $g_{\bar t \bar t}>0$. \label{fig:GPS_ergo}}
\end{minipage}
\end{center}
\end{figure}
\fi

\medskip
{\sl Squashed Kerr-G\"odel black hole solution}:\label{sec:solution}
Now we present the squashed Kerr-G\"odel black hole solution~\cite{TIMN} to Eqs. (\ref{Eineq})-(\ref{Maxeq}), which describes a rotating black hole in the rotating GPS monopole background. 
The metric and gauge potential are given, respectively, by 

\begin{eqnarray}
ds^2=-fdt^2-2g\sigma^3dt+h(\sigma^3)^2+\frac{k^2}{V}dr^2+\frac{r^2}{4}[k\{(\sigma^1)^2+(\sigma^2)^2\}+(\sigma^3)^2],
\end{eqnarray}
and
\begin{eqnarray}
A=\frac{\sqrt{3}}{2}jr^2\sigma^3,
\end{eqnarray}
where the functions $f,g,h,V,k$ in the metric are 
\begin{eqnarray}
&&f(r)=1-\frac{2m}{r^2},\\
&&g(r)=jr^2+\frac{ma}{r^2},\\
&&h(r)=-j^2r^2(r^2+2m)+\frac{ma^2}{2r^2},\\
&&V(r)=1-\frac{2m}{r^2}+\frac{8jm(a+2jm)}{r^2}+\frac{2ma^2}{r^4},\\
&&k(r)=\frac{V(r_\infty)r_\infty^4}{(r^2-r_\infty^2)^2}.
\end{eqnarray}
Taking the limit of $r_\infty\to\infty$ with the other parameters fixed recovers the Kerr-G\"odel black hole solution~\cite{Gimon-Hashimoto} which 
has CTCs in the far region from the horizon. 
For the squashed solution, the parameters $(m,q,a,j,r_\infty)$ should satisfy the inequalities
\begin{eqnarray}
&&m>0,\label{eq:para1}\\
&&\frac{r^2_\infty}{m}>1-4j(a+2jm)>\sqrt{\frac{2}{m}}|a|,\label{eq:para2}\\
&&r_\infty^4-2m(1-4j(a+2jm))r^2_\infty+2ma^2>0,\label{eq:para3}\\
&&-4j^2r^6_\infty+(1-8j^2m)r_\infty^4+2ma^2>0,\label{eq:para4}
\end{eqnarray}
which come from the requirements that there should be two horizons (inner and outer horizons) and there should be no CTC outside the horizons. 
The spacetime has the timelike Killing vector field $\partial_t$ and two spatial Killing vector fields with closed orbits, $\partial_\phi$ and $\partial_\psi$.
Note here that in the limit of $r_\infty\to\infty$ with the other parameters finite, Eq. (\ref{eq:para4}) cannot be satisfied. 
This is the reason why 
one can obtain a Kaluza-Klein black hole solution without CTCs 
by performing the squashing transformation for the Kerr-G\"odel black hole solution with CTCs in the exterior region of the horizon.
In Ref.~\citen{TIMN}, readers can find the squashed Kerr-Newman-G\"odel black hole solution which is a further generalization of the above solution to the charged case, and is specified by five parameters $(m,a,j,q,r_\infty)$. In this review, we do not present the detail of the most general solution, but the squashed Reissner-Nordstr\"om which has the parameters $(m,j,q,r_\infty)$ was studied in Ref.~\citen{TI}.

\medskip
The most remarkable point is that the squashed Kerr-G\"odel black hole solution has two independent rotation parameters $a$ and $j$, where the parameters $a$ and $j$ are related to the rotation of a black hole along the Killing vector $\partial_\psi$ and the rotation of the background ( the squashed G\"odel universe, i.e., the rotating GPS monopole) along the Killing vector $-\partial_\psi$ , respectively. This is why we call these parameters $a$, $j$ {\it Kerr parameter} and {\it G\"odel parameter}, respectively. 
The Komar angular momentum $J_\psi(r)$ over the $r\ (r<r_\infty)$ constant surface is given by
\begin{eqnarray}
J_{\psi}(r)=-\frac{\pi}{2}[2a^2jm+2j^3r^6+m(-1+2j^2(4m+3r^2))a].
\end{eqnarray}
If the black hole and the universe are mutually rotating in the inverse directions, the effect of their rotations can cancel out.  In fact, in spite of the fact that the solution is stationary and non-static, we can see that the angular momentum $J_{\psi}$ with respect to the Killing vector $\partial_\psi$ can vanish at infinity if the parameters $(m,a,j,r_\infty)$ satisfy
\begin{eqnarray}
2a^2jm+2j^3r_\infty^6+m(-1+2j^2(4m+3r_\infty^2))a=0.
\end{eqnarray}
For allowed choices of parameters, all cases 
\begin{eqnarray}
 J_\psi(R_\infty)\lessgtr 0,\ J_\psi(R_+) \lessgtr 0,
\end{eqnarray}
are possible. See Ref. \citen{TIMN} for the detail.

\medskip
{\it Ergoregion}: 
It is interesting to study the structure of ergoregions for the squashed Kerr-G\"odel solution. 
 The metric can be rewritten in the form
\begin{eqnarray}
	ds^2&=&\left(h(r)+\frac{r^2}{4}\right)
		\left[\sigma^3+\left(-\frac{g(r)}{h(r)+\frac{r^2}{4}}C
	+D\right)d\bar t\right]^2\nonumber\\
&&	-\frac{\frac{r^2}{4}V(r)}{h(r)+\frac{r^2}{4}}d{\bar t}^2+\frac{k(r)^2}{V(r)}dr^2+\frac{r^2}{4}k(r)\{(\sigma^1)^2+(\sigma^2)^2\}.
\end{eqnarray}
From the above form, we can easily see that at the outer horizon $r=r_+$ and the infinity $r=r_\infty$, the $\bar t\bar t$-component of the metric takes a non-negative and a negative-definite form:
\begin{eqnarray}
&&g_{\bar t\bar t}(r=r_{+})=\left(-\frac{g(r_{+})}{h(r_{+})+r_{+}^2/4}C+D\right)^2\ge 0,\\
&&g_{\bar t\bar t}(r=r_\infty)=-\frac{r^2_\infty}{4}\frac{V(r_\infty)}{h(r_\infty)+r_\infty^2/4}C^2<0.
\end{eqnarray}
Hence, as we have mentioned previously, the ergosurfaces are located at $r$ such that $g_{\bar t\bar t}=0$, i.e., the roots of the cubic equation with respect to $r^2$, $F(r^2):=g_{\bar t\bar t}r^4=0$. 
The black hole spacetime admits the existence of two ergoregions in the domain of outer communication, just around the horizon and in the far region from the black hole. 
These two ergoregions can rotate in the opposite direction as well
as in the same direction~\cite{TIMN}.

\subsubsection{Multi-black holes in the rotating GPS monopole background}

The squashed Kerr-Newman-G\"odel black hole solution above was also generalized to multi-black hole solution~\cite{MINT}, which 
 is seen to be also obtained by the method~\cite{Gauntlett}.
The forms of the metric and the gauge potential one-form are
\begin{eqnarray}
 ds^2 &=& - H^{-2} \left[ dt + \alpha V  \left( dx^5 +\bm \omega \right) \right] ^2 + H ds^2_{\rm TN},  \label{metric} \\
  A &=&  \frac{\sqrt 3}{2} H^{-1} \left[ dt + \alpha V  \left( dx^5 +\bm \omega \right) \right],  \label{gauge}      
\end{eqnarray}  
where the function $H$ and the metric $ds^2_{\rm TN} $ are given by
\begin{eqnarray}             
 H &=& 1 + \sum _i \frac{M_i}{\left| \bm R - \bm R _i \right|} ,  \label{H} \\
ds^2_{\rm TN} &=& V^{-1} ds_{{\mathbb E}^3}^2 
                  + V \left( dx^5 +\bm \omega \right) ^2 ,   \label{GHmetric}\\
                 V^{-1} &=& 1 + \sum _i \frac{N_i}{\left| \bm R - \bm R _i \right|}, \label{nut}
\end{eqnarray}
respectively,
where $ds_{{\mathbb E}^3}^2=dx^2+dy^2+dz^2 $ is a metric on the three-dimensional Euclid space, ${\mathbb E}^3$, and ${\bm R}=(x,y,z)$ denotes a position vector on ${\mathbb E}^3$. The function $V^{-1}$ is a harmonic function 
on ${\mathbb E}^3$ with point sources located at 
$\bm R=\bm R_i:=(x_i,y_i,z_i)$, where the Killing vector field $\partial_{x^5}$ has fixed points in the base space. 
The one-form $\bm \omega$, which is determined by
\begin{eqnarray}
\bm \nabla \times {\bm \omega}=\bm \nabla V^{-1}, 
\end{eqnarray}
has the explicit form
\begin{eqnarray}                 
 \bm \omega &=& \sum_{i} 
   N_i ~ \frac{z-z_i}{\left| \bm{R}-\bm{R}_i \right|} ~ 
   \frac{(x-x_i) dy -(y-y_i) dx}{(x-x_i)^2+(y-y_i)^2}, \label{omega}
\end{eqnarray}
with $M_i,N_i$ and $\alpha$ being constants.
The parameter region is give by the following inequalities
\begin{eqnarray}
M_i>0,\quad N_i>0 , \quad 0 \le \alpha^2 < 1 ,
\end{eqnarray}
which is necessary and sufficient conditions for the absence of singularity and closed time like curve in the exterior region of black holes.

\medskip

A black hole horizon exists at the position of the harmonic function $H$ and $V^{-1}$, i.e.,  $R=0$. 
In the coordinate system $(t ,R ,\theta , \phi ,x^5)$, 
the metric \eqref{metric} diverges at $R=0$, however this is apparent. 
In order to remove this apparent divergence, 
we introduce a new coordinate $v$ such that 
\begin{eqnarray} 
 dv = dt - \sqrt{\left( 1 + \frac{M}{R} \right) ^3 \left( 1 + \frac{N}{R} \right) } dR . 
\end{eqnarray}
Then, near $R \simeq 0$, the metric \eqref{metric} behaves as 
\begin{eqnarray} 
 ds^2 \simeq -2 \sqrt{\frac{N}{M}} dvdR 
+ MN \left[ d\Omega_{S^2} ^2 + \left( \frac{d x^5 }{N} + \cos \theta d\phi \right) ^2 \right]+{\cal O}(R).    \label{metrich}
\end{eqnarray}
This metric well behaves at $R=0$. The Killing vector field $V =\partial_v$ becomes null at $R=0$ and $V$ is hypersurface orthogonal from $V_\mu dx^\mu = g_{vR} dR$ at the place. 
Therefore the hypersurface, $R = 0$, is a Killing horizon.  
In the coordinate system $(v ,R ,\theta,\phi,x^5)$, 
each component of the metric is analytic in the region of $R\ge 0$. Hence the   space-time has no curvature singularity on and outside the black hole horizon. 
The induced metric on the three-dimensional 
spatial cross section of the black hole horizon located at $R=0$ with the timeslice is obtained as  
\begin{eqnarray}
 \left. ds^2 \right| _{R=0,v={\rm const.}} 
 =\frac{LMn}{2} \left[ d\Omega _{S^2} ^2 + \left( \frac{d\psi}{n} + \cos \theta d\phi \right) ^2 \right] 
 = 2 M L n d\Omega _{\rm S^3/{\mathbb Z}_n} ^2,       
\end{eqnarray}
where $d\Omega _{\rm S^3/{\mathbb Z}_n} ^2$ denotes the metric on the lens space $L(n;1)={S^3}/{\mathbb Z}_n$ with a unit radius. In particular, in the case of $n=1$, the shape of the horizon is a round $S^3$ in contrast to the non-zero angular momentum case \cite{IM,IKMT}.

\subsection{Other Kaluza-Klein black hole solutions}

\subsubsection{Caged black holes}
Myers constructed a {\it caged black hole} as a supersymmetric solution in the five-dimensional minimal supergravity~\cite{myers}. First, he constructed the five-dimensional dimensional Majumdar-Papapetrou multi-black hole solution, 
which can be obtained by replacing the Gibbons-Hawking space metric 
with the flat space metric $ds^2=dx^2+dy^2+dz^2+dw^2$. 
Next,  by superimposing an infinite number of black holes aligned in $w$ direction with an equal separation $a$ in the five-dimensional Majumdar-Papapetrou space-time, he compactified the space-time with the period $2\pi a$ in the direction. The resulting spacetime depends on the $5$-th coordinate $w$ along the the compactification and hence the black holes are not localized at the nut. 
It is clear that the black hole size in the direction is smaller than the scale of compactification, namely,  the black hole is caged in the extra-dimension. For this reason, this is called a caged black hole.
The harmonic function of the caged black hole can be written
\begin{eqnarray}
H=1+\frac{\pi\mu}{a\rho}\frac{\sinh\pi\frac{\rho}{a}\cosh\pi\frac{\rho}{a}}{\sin^2\pi\frac{w}{a}+\sinh^2\pi\frac{\rho}{a}},
\end{eqnarray}
where $\rho=x^2+y^2+z^2$. This describes Kaluza-Klein black holes which asymptote to the direct product 
of the four-dimensional Minkowski space-time and $S^1$.
The further generalizations to higher dimensions~\cite{myers}, or the rotational case~\cite{MOT} were considered.

\subsubsection{Sequences of bubbles and black holes}
As is well known, one of the most important features in Kaluza-Klein theory is that it admits the existence of Kaluza-Klein {\it bubble of nothing}~\cite{Witten-bubble}, 
where the $5$-th dimensional direction smoothly shrinks to zero at finite radius and hence the spacetime is regular. 
In particular, the Kaluza-Klein bubbles not only reveal non-uniqueness property for black holes in Kaluza-Klein theory but also play an important roles in the configuration of black holes, namely, they can  
 keep black holes in static equilibrium because they can balance black holes against the gravitational attraction between them. 
The exact solutions describing black holes on a bubble were first found by Emparan and Reall within a class of the generalized Weyl solutions~\cite{Weyl}.
The solutions describing two black holes held apart by a bubble were studied~\cite{Elvang-H,TIM,IMT}.
Furthermore, for five and six-dimensions, the static general sequences of bubbles and black holes were studied~\cite{EHO} (See Ref.~\citen{Harmark-Obers} for phases of Kaluza-Klein black holes with bubble).
Recently, the various  generalizations to the (magnetic) charged cases~\cite{Yaza,Yaza2} and black ring case~\cite{Yaza3} have been made.

\section*{Acknowledgements}
We would like to thank Ken Matsuno, Masashi Kimura, Toshiharu Nakagawa and Akihiro Ishibashi for useful discussions.
We also thank Maria J. Rodriguez for the discussion concerning the parameter region of the Pomeransky-Sen'kov black ring solutions.
ST is supported by the JSPS under Contract No.20-10616.
HI is supported by Grant-in-Aid for Scientific Research No.19540305.


\begin{thebibliography}{99}

\bibitem{Sch}
K. Schwarzschild, {\it Sitzber. Deut. Akad. Wiss. Berlin}, KL. Math.-Phys. Tech. (1916), 189.

\bibitem{Kerr}
R. P. Kerr, Phys. Rev. Lett. {\bf 11} (1963), 237. 

\bibitem{ER-review}
R. Emparan and H. S. Reall,  Living Rev. Rel. {\bf 11} (2008), 6.
\bibitem{ER-review2}
R. Emparan and H. S. Reall, Class. Quant. Grav. {\bf 23} (2006), R169.

\bibitem{MishimaIguchi}
T.~Mishima and H.~Iguchi, Phys. Rev. D {\bf 73} (2006), 044030.
\bibitem{Fig}
P. Figueras, JHEP {\bf 0507} (2005), 039. 
\bibitem{Evslin}
J. Evslin, JHEP {\bf 0809} (2008), 004.
\bibitem{CT}
Y. Chen and E. Teo, Nucl. Phys. B {\bf 838} (2010), 207.

\bibitem{Elvang-Fig}
H. Elvang and P. Figueras, JHEP {\bf 0705} (2007), 050.
\bibitem{diring} 
H. Iguchi and T. Mishima, Phys. Rev. D {\bf 75} (2007), 064018.
\bibitem{Elvang-R}
H. Elvang and M. J. Rodriguez, JHEP {\bf 0804} (2008), 045.
\bibitem{Izumi}
K. Izumi, Prog. Theor. Phys. {\bf 119} (2008), 757.
\bibitem{TanTeo}
H.S. Tan and E. Teo, Phys. Rev. D {\bf 68} (2003), 044021.
\bibitem{multi-MP}
C. A. R. Herdeiro, C. Rebelo, M. Zilhao and M. S. Costa, JHEP {\bf 0807} (2008), 009.



\bibitem{Myers:1986un}
R.~C.~Myers and M.~J.~Perry,
Annals Phys.\  {\bf 172} (1986), 304.
\bibitem{Tangherlini}
F. R. Tangherlini, Nuovo Cimento {\bf 27} (1963), 636.





\bibitem{Weyl0}
H. Weyl, Ann. Phys. ~(Leipzig) {\bf 54} (1917), 117.
\bibitem{Papapetrou1}
A. Papapetrou, Ann. Phys. (Berlin) {\bf 12} (1953), 309.
\bibitem{Papapetrou2}
A. Papapetrou, Ann. Inst. Henri Poincar\'e, A {\bf 4} (1966), 83.
\bibitem{Weyl}
R. Emparan and H. S. Reall, Phys. Rev. D {\bf 65} (2002), 084025.
\bibitem{Harmark}
T. Harmark, Phys. Rev. D {\bf 70} (2004), 124002.
\bibitem{Hollands}
S. Hollands and S. Yazadjiev, Commun. Math. Phys. {\bf 283} (2008), 749.



\bibitem{Cai}
M. I. Cai and G. J. Galloway, Class. Quant. Grav. {\bf 18} (2001), 2707.
\bibitem{Helfgott}
C. Helfgott, Y. Oz and  Y. Yanay, JHEP {\bf 0602} (2006), 025.
\bibitem{galloway}
G. J. Galloway and R. Schoen, Commun. Math. Phys. {\bf 266} (2006), 571. 







\bibitem{Emparan:2001wn}
R.~Emparan and H.~S.~Reall,
Phys.\ Rev.\ Lett.\  {\bf 88} (2002), 101101.
\bibitem{Pom}
A. A. Pomeransky and R. A. Sen'kov, e-Print: arXiv:hep-th/0612005.


\bibitem{Gauntlett}
J. P. Gauntlett, J.B. Gutowski, C.M. Hull, S. Pakis and H.S. Reall, 
Class. Quant. Grav. {\bf 20} (2003), 4587. 
\bibitem{IKMT}
H. Ishihara, M. Kimura, K. Matsuno and S. Tomizawa, Class. Quant. Grav. {\bf 23} (2006), 6919.



\bibitem{shiromizu}
G. W. Gibbons, D. Ida and T. Shiromizu,
Prog. Theor. Phys. Suppl. {\bf 148} (2002), 284;\\
G. W. Gibbons, D. Ida and T. Shiromizu,
Phys. Rev. Lett. {\bf 89} (2002), 041101.


\bibitem{uniqueness}
For review, M.~Heusler,
{\it Black Hole Uniqueness Theorems},
(Cambridge University Press, Cambridge, 1996).
\bibitem{Israel}
W. Israel, Phys. Rev. {\bf 164} (1967), 1776.
\bibitem{Bunting}
G. L. Bunting and A. K. M. Masood-ul-Alam, Gen. Rel. Grav. {\bf 19} (1987), 147.





\bibitem{Emparan2}
R. Emparan, JHEP {\bf 03} (2004), 064.



\bibitem{TMY}
S. Tomizawa, Y. Morisawa and Y. Yasui, Phys. Rev. D {\bf 73} (2006), 064009.





\bibitem{MTY}
Y. Morisawa, S. Tomizawa and Y. Yasui, Phys. Rev. D {\bf 77} (2008), 064019. 

\bibitem{KN}
D. Kramer and G. Neugebauer, Phys. Lett. A {\bf 75} (1980), 259.

\bibitem{Hoen}
C. Hoenselaers, Prog. Theor. Phys. {\bf 72} (1984), 761.



\bibitem{DH}
W. Dietz and C. Hoenselaers, Ann. Phys. (N.Y.) {\bf 165} (1985), 319.


\bibitem{eact}
H. Stephani, D. Kramer, M. MacCallum, C. Hoenselaers and E. Herlt, {\it Exact Solutions to
Einstein's Field Equations Second Edition}, Cambridge University Press, Cambridge (2003).

\bibitem{NH}
G. Neugebauer and J. Hennig, Gen. Rel. Grav. {\bf 41} (2009), 2113. 


\bibitem{Wald}
R. Wald,  Phys. Rev. D {\bf 6} (1972), 406.
























































\bibitem{Evslin-Krishnan}  
J.~Evslin and C.~Krishnan, Class. Quant. Grav. {\bf 26} (2009), 125018.  
  



\bibitem{thermo-diring}
H. Iguchi and T. Mishima,  Phys. Rev. D {\bf 82} (2010), 084009.



\bibitem{private}
We would like to thank Iguchi and Mishima for informing us of typos and discussing the parameter counting.







\bibitem{IM}
H. Ishihara and K. Matsuno, Prog. Theor. Phys. {\bf 116} (2006), 417.
\bibitem{Wang}
T. Wang, Nucl. Phys. B {\bf 756} (2006), 86. 
\bibitem{NIMT}
T. Nakagawa, H. Ishihara, K. Matsuno and S. Tomizawa, Phys. Rev. D {\bf 77} (2008), 044040. 
\bibitem{TIMN}
S. Tomizawa, H. Ishihara, K. Matsuno and T. Nakagawa, Prog. Theor. Phys. {\bf 121} (2009), 823.
\bibitem{MINT}
K. Matsuno, H. Ishihara, T. Nakagawa and S. Tomizawa, Phys. Rev. D {\bf 78} (2008), 064016. 
\bibitem{TI}
S. Tomizawa and A. Ishibashi, Class. Quant. Grav. {\bf 25} (2008), 245007.
\bibitem{T}
S. Tomizawa, e-Print: arXiv:1009.3568 [hep-th].





\bibitem{Gaiotto}
D. Gaiotto, A. Strominger and X. Yin, JHEP {\bf 02} (2006), 023.
\bibitem{Elvang3}
H. Elvang, R. Emparan, D. Mateos and H. S. Reall, JHEP {\bf 0508} (2005), 042.




\bibitem{LMP}
H. Lu, Jianwei Mei and C.N. Pope, Nucl. Phys. B {\bf 806} (2009), 436.
\bibitem{LMP2}
H. Lu, Jianwei Mei and C.N. Pope, Class. Quant. Grav. {\bf 27} (2010), 075013. 


\bibitem{TYM}
S. Tomizawa, Y. Yasui and Y. Morisawa, Class. Quant. Grav. {\bf 26} (2009), 145006.
\bibitem{GS}
D. V. Gal'tsov and N. G. Scherbluk, Phys. Rev. D {\bf 79} (2009), 064020.





\bibitem{Rasheed}
D. Rasheed, Nucl. Phys. B {\bf 454} (1995), 379. 



  
\bibitem{Chodos-Detweiler}
A. Chodos and S. Detweiler, Gen. Rel. Grav. {\bf 14} (1982), 879.
\bibitem{FZB}
V. P. Frolov, A. I. Zel'nikov and U. Bleyer, Ann. Physik, {\bf 44} (1987), 371.
\bibitem{DM}
P. Dobiasch and D. Maison, Gen. Rel. Grav. {\bf 14} (1982), 231. 
\bibitem{GW}
G. W. Gibbons and D. L. Wiltshire, Ann. Phys. {\bf 167} (1986), 201.
\bibitem{Pollard}
D. Pollard, J. Phys. A {\bf 16} (1983), 565.

\bibitem{Gibbons-Maeda}
G. W. Gibbons and K. Maeda, Nucl. Phys. B {\bf 298} (1988), 741.



\bibitem{Maison}
D. Maison, Gen. Relativ. {\bf 10} (1979), 717.




\bibitem{CY-KK}
M. Cveti{\v c} and D. Youm, Phys. Rev. D {\bf 52} (1995), 2144.


\bibitem{CY-KK2}
M. Cveti{\v c} and D. Youm, Phys. Rev. Lett. {\bf 75} (1995), 4165.




\bibitem{Bena}
I. Bena and P. Kraus, Phys. Rev. D {\bf 70} (2004), 046003. 
\bibitem{Bena2}
I. Bena, P. Kraus and R. Warner, Phys. Rev. D {\bf 72} (2005), 084019.
\bibitem{Bena3}
I. Bena and N. P. Warner, Adv. Theor. Math. Phys. {\bf 9} (2005), 667.
\bibitem{BGRW}
I. Bena, S. Giusto, C. Ruef and N. P. Warner, JHEP {\bf 0911} (2009), 032.
\bibitem{BKW}
I. Bena, P. Kraus and N.P. Warner, Phys. Rev. D {\bf 72} (2005), 084019. 



\bibitem{EEMR2}
H. Elvang, R. Emparan, D. Mateos and H. S. Reall, Phys. Rev. D {\bf 71} (2005), 024033.



\bibitem{FGPS}
J. Ford, S. Giusto, A. Peet and A. Saxena, Class. Quant. Grav. {\bf 25} (2008), 075014.
\bibitem{CEFGS}
J. Camps, R. Emparan, P. Figueras,  S. Giusto and A. Saxena, JHEP {\bf 0902} (2009), 021.
\bibitem{GRS}
S. Giusto,  S. F. Ross and  A. Saxena, JHEP {\bf 0712} (2007), 065. 
\bibitem{BDGRW}
I. Bena, G. Dall'Agata,  S. Giusto, C. Ruef and N. P. Warner, JHEP {\bf 0906} (2009), 015.
\bibitem{CBJV}
G. Compere, S. Buyl, E. Jamsin and A. Virmani, Class. Quant. Grav. {\bf 26} (2009), 125016.
\bibitem{Gibbons-Perry}
G. W. Gibbons and M. J. Perry, Nucl. Phys. B {\bf 248} (1984), 629.
\bibitem{CY-KK3}
M. Cveti{\v c} and D. Youm, Nucl. Phys. B {\bf 438} (1995), 182.
\bibitem{CY-KK4}
M. Cveti{\v c} and D. Youm, Nucl. Phys. B {\bf 453} (1995), 259.
\bibitem{CY-KK5}
M. Cveti{\v c} and D. Youm, Phys. Rev. D {\bf 52} (1995), 2574.
\bibitem{Nelson}
W. Nelson, Phys. Rev. D {\bf 49} (1994), 5302. 






\bibitem{CY96}
M. Cveti\v{c} and D. Youm, Nucl. Phys. B {\bf 476} (1996), 118.




\bibitem{Gimon-Hashimoto}
E. Gimon and A. Hashimoto, Phys. Rev. Lett. {\bf 91} (2003), 021601.
\bibitem{Herdeiro}
C. A. R. Herdeiro, Class. Quant. Grav. {\bf 20} (2003), 4891.
\bibitem{Wu}
S-Q. Wu, Phys. Rev. Lett. {\bf 100} (2008), 121301.







\bibitem{Galtsov}
A. Bouchareb, G. Cl\'ement, C-M. Chen, D. V. Gal'tsov, N. G. Scherbluk, and T. Wolf, Phys. Rev.
D {\bf 76}, 104032 (2007); Erratum-ibid. D {\bf 78} (2008), 029901.
\bibitem{Galtsov2}
D. V. Gal'tsov and N. G. Scherbluk, Phys. Rev. D {\bf 78} (2008), 064033.




\bibitem{GH}
G. W. Gibbons and S.W. Hawking, Phys. Lett. B {\bf 78} (1978), 430.


\bibitem{GP}
D. J. Gross and M. J. Perry, Nucl. Phys. B {\bf 226} (1983), 29; \\
R. D. Sorkin, Phys. Rev. Lett. {\bf 51} (1983), 87. 


\bibitem{myers}
R. C. Myers, Phys. Rev. D {\bf 35} (1987), 455.



\bibitem{MOT}
K. Maeda, N. Ohta and M. Tanabe, Phys. Rev. D {\bf 74} (2006), 104002.

%<<<<<<<<<<<<<<<<<<< D=5 Supersymmetric solutions     >>>>>>>>>>>>>>>>>>>>>>>>>%









\bibitem{Witten-bubble}
E. Witten, Nucl. Phys. B {\bf 195} (1982), 481.
\bibitem{Elvang-H}
H. Elvang and G. T. Horowitz, Phys. Rev. D {\bf 67} (2003), 044015.
\bibitem{TIM}
S. Tomizawa,  H. Iguchi and T. Mishima, Phys. Rev. D {\bf 78} (2008), 084001.
\bibitem{IMT}
H. Iguchi, T. Mishima and S. Tomizawa, Phys. Rev. D {\bf 76} (2007), 124019; Erratum-ibid. Phys. Rev. D {\bf 78} (2008), 109903.
\bibitem{EHO}
H. Elvang, T. Harmark and N. A. Obers, JHEP, {\bf 0501} (2005), 003.
\bibitem{Harmark-Obers}
T. Harmark and N. A. Obers, e-Print: arXive: hep-th/0503020.


\bibitem{Yaza}
J. Kunz, and S. S. Yazadjiev,  Phys. Rev. D {\bf 79} (2009), 024010.


\bibitem{Yaza2}
S. S. Yazadjiev and P. G. Nedkova, JHEP {\bf 1001} (2010), 048.


\bibitem{Yaza3}
P. G. Nedkova and S. S. Yazadjiev, Phys. Rev. D {\bf 82} (2010), 044010.


\end{thebibliography}
\end{document}